\documentclass[11pt,letterpaper]{article}
\usepackage{jheppub}
\usepackage{bm}

\newcommand{\N}{\mathcal{N}}
\renewcommand{\t}{\tilde}
\newcommand{\del}{\partial}
\newcommand{\Del}{\nabla}
\newcommand{\comment}[1]{}
\newcommand{\beq}[1]{\begin{equation}\label{#1}}
\newcommand{\eeq}{\end{equation}}
\newcommand{\bea}{\begin{eqnarray}}
\newcommand{\eea}{\end{eqnarray}}
\newcommand{\h}{\hat}
\newcommand{\B}{\bm{B}}
\newcommand{\w}{\wedge}
\renewcommand{\L}{\mathcal{L}}
\newcommand{\bG}{\bar{G}_3^{(0)}}
\newcommand{\G}{G_3^{(0)}}
\newcommand{\tF}{\tilde{F}}
\renewcommand{\b}{\bar}
\newcommand{\0}{(0)}

\title{The Dimensional Reduction and K\"ahler Metric of Forms In Flux and Warping}

\author{Andrew R. Frey}
\author{and James Roberts}
\affiliation{Department of Physics and Winnipeg Institute for Theoretical 
Physics, University of Winnipeg, Winnipeg, Manitoba, Canada R3B 2E9}

\emailAdd{a.frey@uwinnipeg.ca}

\abstract{We present a first-principles derivation of the K\"ahler metric
for axion-like moduli of conformally Calabi-Yau compactifications of IIB
string theory with imaginary self-dual 3-form flux at the classical level.  
We find that the warp factor
and flux modify the moduli space metric and therefore K\"ahler potential even
in classical supergravity, with the modifications scaling as (volume)$^{-2/3}$ in
the large-volume limit. Our derivation
emphasizes the role of constraints from 10D gauge symmetries and highlights
metric formality as a geometric property that protects the moduli space
of highly supersymmetric toroidal orientifolds.  Our results have important
quantitative implications for nonperturbative moduli stabilization, 
phenomenology, and cosmology in flux compactifications.}%, even in the large-volume

\begin{document}
\maketitle

\section{Introduction}\label{intro}

The twenty-first century has seen the rapid development of string theory 
compactifications beyond the traditional product of four-dimensional (4D)
Minkowski spacetime with a Calabi-Yau (CY) 3-fold.  The key ingredient
has been the addition of flux in the metric and supergravity form fields,
which allows the construction of a wide variety of ten-dimensional (10D)
backgrounds (extending a relatively smaller literature from before the turn
of the century), which often include warping (see, for example, 
\cite{hep-th/0308156,hep-th/0509003,arXiv:0708.3984} for reviews).  
Great attention has been paid to the role of fluxes
in stabilizing moduli at the classical level.  
Furthermore, this variety of backgrounds has
provided a framework for exploring phenomenology and cosmology in string 
theory, including the embedding of inflation in 10D theories.

However, our understanding of dimensional reduction around these backgrounds,
even at the level of understanding moduli stabilization, is not so well
developed.  Most of the work on moduli stabilization and cosmology is 
from the perspective of 4D effective potentials heuristically associated
with 10D physics, and it is usually assumed
that the space of scalar fields can be described as the moduli space of a 
CY compactification with a scalar potential added.  This assumption extends
to assuming that even unprotected quantities such as the K\"ahler potential
remain unmodified by the flux and warping.  Since these backgrounds typically
have a low degree of (or no) supersymmetry, the compactification
geometry is often not CY, and the mass scale of moduli stabilization can be 
as high as the Kaluza-Klein scale, it should be clear that a more 
fundamentally 10D approach is needed when determining what 4D effective 
theory arises in a given 10D background.  

From a 10D point of view, 
the moduli can be identified with 
spacetime-independent deformations that lead to a background which still
solves the 10D supergravity equations of motion (EOM)
(or leaves the string worldsheet theory conformally invariant, when a 
worldsheet description is feasible).  Finding other aspects
of the effective theory, such as the metric on moduli space, requires
identifying the zero-modes of small fluctuations around the 10D background
(\textit{ie}, those fluctuations that satisfy massless wave equations in 4D
when on shell).
Unfortunately, in nontrivial backgrounds, the 10D gauge and diffeomorphism
complicate this process, specifically leading to nontrivial constraints
\cite{hep-th/0201029,hep-th/0308156,hep-th/0507158,arXiv:0803.3068}
and the gauge identification of apparently distinct degrees of freedom
\cite{arXiv:1009.4200}.  Failing to deal with these issues properly can lead 
to misleading results; \cite{arXiv:0805.3700,arXiv:1009.4200} have 
presented formalisms to elucidate the physical meaning of the constraints.
One of the goals of this paper is to provide experience in solving
constraints, which will hopefully be applicable to a broad class of flux
compactifications. 

In this paper, we present a solution to the constraints and a first-principles
derivation of the moduli space metric for axionic moduli of conformally
CY compactifications with warping and imaginary self-dual (ISD) 3-form flux
in type IIB string theory, as first discussed in 
\cite{hep-th/9908088,hep-th/0004103,hep-th/0009211,hep-th/0010010,hep-th/0105097} 
(following the work of
\cite{hep-th/9605053} in M theory).  These compactifications possess a 
no-scale structure, a volume modulus at the classical level, and either 
unbroken or spontaneously broken supersymmetry.  As a result, the metric on
moduli space is controlled by a K\"ahler potential, and we find the K\"ahler
potential in one simplifying case as well as comment on the K\"ahler
potential of all closed string moduli.  While the no-scale structure 
protects the K\"ahler potential from corrections (up to field redefinition)
in the case of a single modulus, the form of the K\"ahler potential
is considerably less restricted
for multiple moduli \cite{arXiv:1205.5728}, so we expect to 
find corrections due to the presence of flux and warping.  In fact,
\textit{we demonstrate that generally flux and warping do correct the metric
on moduli space} and give an explicit geometric condition under which the 
K\"ahler metric is uncorrected.  This condition is consistent with the fact
that the moduli space of $\N=3$ and $\N=4$ compactifications on warped tori
is determined by supersymmetry 
\cite{deRoo:1984gd,Bergshoeff:1985ms,Castellani:1985ka}.  
\comment{It is also worth noting that the corrections
due to the flux do not disappear at large volume, even though the
warp factor becomes trivial in that limit.}
It is worth noting that the corrections we find are distinct from those discussed
in \cite{Becker:2002nn,Grimm:2013gma,Pedro:2013qga}, 
which originate in higher-curvature terms %on D7-branes and O7-planes 
and were derived ignoring the warp factor, whereas
we consider only classical supergravity but include the effects of nontrivial
warping and flux. 
Our discussion extends the results of
\cite{arXiv:0810.5768}, which found the K\"ahler potential for
the universal volume modulus and its axionic partner.  While we restrict here 
to classical moduli, comments on flux-stabilized massive scalars were 
presented in \cite{hep-th/0603233}. 

The fact that the K\"ahler potential is no longer completely specified by the
topology of the underlying CY manifold has important physical consequences
in string phenomenology and cosmology.
For example, the moduli we study can develop nonperturbative 
superpotentials and therefore serve as inflatons in several embeddings
of string inflation, including racetrack inflation 
\cite{hep-th/0406230,hep-th/0603129}, N-flation \cite{hep-th/0507205},
roulette inflation \cite{hep-th/0612197},
and monodromy inflation \cite{arXiv:0808.0706} (the perturbative potential
for axions coupled to branes in monodromy inflation should be calculated
using correct solutions to the constraints, as well).  Since the K\"ahler
potential contributes to the scalar potential along with the superpotential,
the corrections we identify potentially affect all the cosmological observables
controlled by the inflationary potential, including which regions of 
field space satisfy the slow-roll conditions.  Similarly, 
\cite{Pedro:2013qga} recently described how other K\"ahler potential corrections 
affect the stabilization of classical moduli; much of their analysis applies
directly to our results, since both sets of corrections to the K\"ahler potential
have the same scaling with compactification volume in the large-volume limit.
In many ways, this is a 
more general statement of cautions presented in 
\cite{hep-th/0307084,hep-th/0312076,hep-th/0407126} about determining 
potentials in dimensional reduction without first finding the proper 10D 
modes corresponding to light but massive 4D degrees of freedom.  

The plan of this paper is as follows.  In the next section, we sketch our
strategy for a first-principles derivation of the moduli space metric in
4D effective theory.  Then, in section \ref{review}, we review the
class of compactifications we study, including their moduli and moduli 
stabilization by 3-form flux.  We also comment on features special to 
internal manifolds with torus factors.
Section \ref{warped} presents the constraint equations and dynamical
EOM for axions descending from 10D 2- and 4-form gauge fields as well as a
review the universal volume modulus from \cite{arXiv:0810.5768}.
The section concludes with a discussion of the mathematical concept of 
metric formality and its importance in determining the solution of the 
constraints.  The following section derives the 4D moduli space metric (and
K\"ahler potential in simple cases), 
again emphasizing the importance of metric formality
and indicating when the flux and warping correct the K\"ahler potential.
Finally, we close with a brief discussion of our results in section 
\ref{discuss}.  Our conventions, useful results, and derivations that fall
outside the main flow of the manuscript appear in the appendices.

\section{Systematics of dimensional reduction}\label{dimred}

Here we briefly present a strategy for a first-principles derivation of a 
lower-dimensional effective theory from a higher-dimensional theory,
specializing at the end to calculation of the K\"ahler metric and potential
in 4D supergravity.  These techniques rely on the higher-dimensional EOM
and are completely independent of supersymmetry and special geometric or 
topological properties of the internal manifold.  While this procedure is
not new (and is essentially identical to basic Kaluza-Klein reduction), 
we emphasize it because indirect techniques based on relations between 
supersymmetry requirements in the different dimensions (as in 
\cite{arXiv:0902.4031}), cohomology, or structure group representations
are not guaranteed to capture all the relevant physics in every background.

Consider a $D$-dimensional spacetime with an $n$-dimensional external
maximally symmetric spacetime (we will specialize to $D=10$ and $n=4$
Minkowski spacetime) with metric
\beq{Ddimmetric}
ds^2 = e^{2A(y)} \h g_{\mu\nu}(x) dx^\mu dx^\nu + g_{mn}(y)dy^m dy^n\ .\eeq
In general, the other fields of the $D$-dimensional theory have nontrivial 
backgrounds.  Then decompose spacetime-dependent 
perturbations of the $D$-dimensional fields around the background by 
separation of variables, as usual.  At this stage, it is important to 
determine which perturbations are equivalent under $D$-dimensional gauge
and diffeomorphism invariance.  As \cite{arXiv:1009.4200} has emphasized,
what may appear to be distinct perturbations can be gauge-identified in
nontrivial backgrounds; the prime example is the identification of the 
dilaton fluctuation with the volume modulus in warped compactifications with
a nontrivial dilaton profile.  To find the distinct $n$-dimensional degrees
of freedom, it is possible either to find gauge-invariant variables (much
like in cosmological perturbation theory) \cite{arXiv:1009.4200} or to
fix a gauge.

Next, expand the $D$-dimensional EOM to 
desired order in perturbation theory (we will see below that linear order
is sufficient to determine kinetic terms and masses).  
The $D$-dimensional EOM for perturbations divide into two classes.  First
are dynamical EOM, which take the form $\Box\phi(x)\psi(y) \sim \phi\triangle\psi$ 
for the $n$-dimensional Laplacian (appropriate to the spin) and some
$D-n$-dimensional operator $\triangle$.  $n$-dimensional mass eigenstates are
eigenfunctions of $\triangle$, and masses are given by the eigenvalues.  
The $n$-dimensional effective field theory is concerned with light
modes, which are typically zero-modes of $\triangle$ in some approximation or
limit.  The other EOM are a constraint equations, which arise due to 
gauge invariance as in any other constrained system.  Constraints are
schematically of the form $\Del_\mu\phi \diamondsuit_m\psi =0$ (for bosons) and 
must be satisfied
at every point in spacetime.  A familiar constraint of this form is the
Gauss law constraint in electrodynamics, which is $\del_t\del_i A_i=0$ in 
temporal gauge in the ``dimensional reduction'' from 4 to 1 dimension.  
Generically, the constraints mix apparently unrelated $D$-dimensional fields
in the perturbations associated with a single $n$-dimensional degree of freedom
(the ``additional'' field components required are known as ``compensators'').

It is worth a brief diversion to appreciate the role of constraints in 
traditional compactifications, such as any string theory compactified with no
sources (included unwarped CY backgrounds).  The zero mode of a 2-form potential
must be closed on the extra dimensions, or else the nontrivial field strength
would cost energy even for a static deformation.  The reader might assume
that a gauge choice allows us to take the profile on the CY to be harmonic;
however, this is not the case.  Specifically, 
$a(x)(\omega_2+\t d\lambda_1)$ transforms
into $a\omega_2 -\h da \lambda_1$ rather than $a\omega_2$ under a 10D gauge
transformation parameterized by $a\lambda_1$.  However, the 10D EOM $d\star dA_2=0$
for $\delta A_2=a(x)\omega_2(y)$ includes a constraint 
$\h\star\h d a (\t d\t\star\omega_2)=0$, which requires $\omega_2$ be harmonic.
This constraint was noted at least as early as \cite{Candelas:1990pi}; the
important point is that constraints are a familiar part of dimensional reduction.

Once the constraints are solved and the dynamical EOM are diagonalized, the
desired light modes, including all appropriate field mixing, can be inserted
into the $D$-dimensional action.  At this point, the constraints should be 
imposed for the consistency of $D$-dimensional gauge and diffeomorphism 
invariance but the dynamical EOM should not since the action is an off-shell 
quantity.  The $n$-dimensional action is obtained by integrating over the 
compact $D-n$ dimensions.

We are interested in the action for moduli in the lower-dimensional
theory, 
\beq{Smoduli} S_{mod} = -\frac{1}{\kappa_n^2} \int d^nx\, G_{ab}(\phi) 
\del_\mu\phi^a \del^{\h\mu}\phi^b\ .\eeq
Treating the moduli $\phi^a$ as perturbations of the $D$-dimensional 
background, this potentially has contributions at all orders in perturbation
theory.  Fortunately, since the background values $\phi^a_{\0}$ are spacetime
independent, the lowest order result contains the entire moduli space
metric $G_{ab}(\phi_{\0})$ as long as the expansion is carried out around 
an arbitrary point $\phi_{\0}^a$ in moduli space.  This is a tremendous 
simplification, since we only need to find the $D$-dimensional action at second
order.  This second order action is
\beq{action2ndorder}
S_{(2)} = \frac 12 \int d^Dx\, \delta\Phi_A \delta E^A\ ,\eeq
where $\delta \Phi_A$ is the first-order $D$-dimensional field including
fluctuations of all moduli and $\delta E^A$ is the dynamical EOM for $\Phi_A$ 
at linear order in all fluctuations ($\delta E^A=0$ on shell). 
This form for the second-order action 
has appeared before in the literature (see \cite{arXiv:0803.3068}), and we
present a derivation for actions with up to two derivatives 
in appendix \ref{linearizedeom} as we are not aware of a
derivation in the literature.  The form of this for type IIB supergravity is also
given in appendix \ref{linearizedeom}.

We close by returning to our example of a 2-form on an unwarped CY background 
in order to emphasize the role of constraints.  Suppose we ignored the constraint
and wrote $\delta A_2=a(x)(\omega_2+\t d\lambda_1)$ with $\omega_2$ harmonic.
The dynamical EOM is then $\delta E_8 =(\h d\h\star\h d a)
\t\star(\omega_2+\t d\lambda_1)$, and the quadratic action is
\beq{actionexample1}S_{(2)} =\frac 12\int a\w\h d\h\star\h d a
\int (\omega_2+\t d\lambda_1)\w \t\star(\omega_2+\t d\lambda_1) 
=\frac 12\int a\w\h d\h\star\h d a
\int \left(\omega_2\w\t\star\omega_2+\t d\lambda_1\w \t\star\t d\lambda_1\right)
\ .\eeq
On the other hand, in a different gauge, $\delta A_2=a(x)\omega_2-\h da \lambda_1$,
but the dynamical EOM, being gauge-invariant, remains the same.  The second
term of $\delta A_2$ in this gauge has a zero wedge product with the dynamical
EOM, however, just by counting external and internal legs, so the quadratic
action is
\beq{actionexample2}S_{(2)} =\frac 12\int a\w\h d\h\star\h d a
\int\omega_2\w\t\star\omega_2\ .\eeq
It is only when we impose the constraint $\lambda_1=0$ that the action is
gauge invariant.

\section{Conformally CY compactifications with ISD flux}\label{review}
Here we review conformally CY backgrounds of IIB string theory, the behavior
of IIB supergravity fields on CY orientifolds and the light fields in such
compactifications, and the stabilization of some moduli by ISD flux.

\subsection{Review of the background}\label{gkp}

The fact that, in some cases, form flux can leave compactification geometry
unaffected up to warping was first discussed in \cite{hep-th/9605053} for M theory
and translated to IIB string theory on CY manifolds by
\cite{hep-th/9908088,hep-th/0004103,hep-th/0009211,hep-th/0010010,hep-th/0105097}.
As \cite{hep-th/0105097} emphasized, 3-form flux in the IIB picture stabilizes
many of the moduli of the underlying CY geometry, and furthermore, these 
compactifications can realize the physics of \cite{hep-ph/9905221} by embedding
the gravity dual of $SU(N)\times SU(N+M)$ gauge theory \cite{hep-th/0007191}
as warped throats.  

These compactifications have background fields given by
\bea
ds^2&=& e^{2\Omega_{\0}(x)} e^{2A_{\0}(y)} \h\eta_{\mu\nu}dx^\mu dx^\nu 
+e^{-2A_{\0}(y)}\t g_{mn}dy^m dy^n\nonumber\\
\tF_5 &=& e^{4\Omega_{\0}}\h\epsilon\w\t de^{4A_{\0}} +\t\star\t d e^{-4A_{\0}} \ ,\quad
\t\star \G=i\G\ .\label{gkpbackground}\eea
Here and in the rest of the manuscript, we use subscript and superscript $\0$ 
to represent background values.  $\t g_{mn}$ is a CY metric; we will choose
a gauge for perturbations such that $\t g_{mn}$ remains Ricci-flat and 
therefore spacetime-independent in the presence of fluctuations.  As a result,
we do not mark it with $\0$.  The factor $e^{2\Omega}$, which we include for
later convenience, converts the 4D effective theory from Jordan frame to Einstein 
frame and is defined by 
\beq{einsteinframe}
e^{2\Omega} \equiv \frac{\int d^6y\,\sqrt{\t g}}{\int d^6y\,\sqrt{\t g}e^{-4A}}
\eeq
at all orders in fluctuations around the background.  This definition follows
from the argument that the replacement $\h \eta_{\mu\nu}\to\h g_{\mu\nu}$ 
gives the 4D metric; it satisfies
the 4D vacuum Einstein equation in the absence of other fluctuations 
\cite{hep-th/0004103,hep-th/0308156} and the Einstein equation with source in
the presence of a null wave of the volume modulus \cite{arXiv:0810.5768}.

The warp factor is given by the Poisson equation 
\beq{warpfactor}
\t\Del^{\t 2} e^{-4A_{\0}} = -\frac{g_s}{2}\left|\G\right|^{\t 2}-\textnormal{local}
\ ,\eeq
where the local terms are D3-brane charges (which may be induced on 
higher-dimension branes or orientifold planes).  Consistency of this equation
in fact demands that the local sources include O3-planes or O7-planes, 
which carry negative D3-brane charge.  While the presence of 7-branes opens
up the possibility of a full description in F theory, we restrict to the
slightly simpler case of the orientifold limit, in which the background 
axio-dilaton is constant on the internal space.  
Then the complex 3-form can be written as $G_3=dA_2$, where $A_2$
is a complex combination of the NSNS and RR 2-form potentials.  Equation
(\ref{warpfactor}) follows from both the Einstein equation and $\tF_5$ EOM
for ansatz (\ref{gkpbackground}); \cite{hep-th/0105097} showed that, assuming
that a BPS-like inequality on local sources is satisfied, then it must be saturated 
and the background given by (\ref{gkpbackground}).  This no-go theorem
disallows anti-D3-branes at the level of IIB supergravity.

Aside from the ISD condition, $\G$ is harmonic.  For supersymmetry of the 
background, $\G$ must additionally be a (2,1) form in the complex geometry of
the CY as well as primitive, $\t J \G=0$, where $\t J$ is the K\"ahler form 
of the CY.  The wedge product of $\t J$ with a harmonic form 
is harmonic, so the fact that there are no harmonic 5-forms on a generic CY means
that $\G$ is automatically primitive.  This can be a nontrivial condition on
compactifications with torus factors, however.

In fact, there can be two other ``invisible'' backgrounds, described in more
detail below: constant shifts of $C_4$ proportional to a harmonic 4-form on the CY 
and constant shifts of $A_2$ proportional to a harmonic 2-form on the 
CY.\footnote{Some additional conditions may be imposed; see below.}
These are closed and do not contribute to the field strengths, but, because
they are harmonic, they are not gauge trivial.

Finally, we note that these backgrounds have a no-scale structure; specifically,
nothing fixes either the CY volume $\t V\equiv\int d^6y\,\sqrt{\t g}$ 
or $e^{2\Omega_{\0}}$.
As it turns out, a constant rescaling of $\t g_{mn}$ changes the metric in 
(\ref{gkpbackground}) only up to a rescaling of the $x^\mu$ coordinates and is
pure gauge (this follows because $e^{2A_{\0}}$ takes the same scaling by
(\ref{warpfactor}), an argument first due to \cite{hep-th/0507158}).  As we review
in section \ref{volume}, the volume modulus shifts $e^{-4A}$ by a function of
the $x^\mu$ only.  As a result, the large-volume limit takes the warp factor 
to a constant.

\subsection{Orientifold projection, moduli, and flux}
\label{cyorientifolds}
As noted above, the compactifications we consider necessarily include O3- or
O7-planes.  Here we describe the parity of IIB supergravity fields under
O3- and O7-plane orientifold projections as well as the moduli of these
compactifications, deformations that take one background of the type described
above into another.  We also mention the other massless modes, which 
can be identified with unbroken symmetries.  These light degrees of freedom
were described
previously in \cite{hep-th/0403067} for generic CY orientifolds.  We also
comment here on moduli stabilization by 3-form flux.  Because of special 
properties of tori, specifically related to nonzero first Betti number, we
comment on compactifications with torus factors in a separate subsection below.

An orientifold projection consists of both a worldsheet parity operation and a
spacetime involution; the internal manifold must therefore have an involution
symmetry.  O3-planes and O7-planes act with the same worldsheet parity.
Under this worldsheet parity operation, the metric $g_{MN}$, axio-dilaton $\tau$,
and 4-form $C_4$ have positive parity, while the 2-forms have negative parity.
As a result, the supergravity modes that survive the orientifold projection
have positive parity under the involution for $g_{MN}$, $\tau$, and $C_4$ and 
negative involution parity for the 2-forms.   The intrinsic 2-form
parity implies that the background flux $\G$ has negative involution parity.  
In the field
decomposition below, it is useful to decompose the cohomology of the CY into
involution parity even and odd pieces indicated by a superscript $\pm$. 

Upon dimensional reduction to 4D, the metric gives rise to a 4D spacetime metric
$\hat g_{\mu\nu}$ and moduli, which are deformations of $\t g_{mn}$ that leave 
the internal manifold Ricci-flat.  At linear order, $\delta\t g_{mn}$ are zero modes
of the Lichnerowicz operator, and they separate into two categories
\cite{Candelas:1990pi}.  First,
the K\"ahler moduli are in one-to-one correspondence with harmonic (1,1) forms,
which are further positive parity under the orientifold involution.  In other words,
the K\"ahler moduli are associated with the cohomology $H_{1,1}^+$.  A change in
the K\"ahler moduli changes the K\"ahler form $\t J$ of the CY without changing the
complex coordinates. On the other hand, complex structure moduli do alter the
complex coordinates of the CY, deforming the holomorphic (3,0) form $\Omega_3$, 
and are
associated with $H_{2,1}^-$.  Due to the lack of isometries on a general CY, there
is no Kaluza-Klein graviphoton.  Furthermore, maintaining the ISD condition on
$\G$ generically stabilizes all the complex structure moduli and $\tau$ while
leaving the K\"ahler moduli massless.  In this paper, we will consider only
the volume modulus; the full solution for the volume modulus was presented
in \cite{arXiv:0810.5768} and is reviewed below.

4D scalar perturbations of $C_4$ are associated with
harmonic 4-forms on the CY which can be described equally well as spacetime 2-forms
and harmonic 2-forms on the CY due to the self-duality of $\tF_5$.
Since they have positive orientifold parity, there are $h_{1,1}^+$ of these
axions.
These are the axionic partners of the K\"ahler moduli of the metric, so 
they remain massless in the presence of flux; their
dimensional reduction is a major focus of this paper.  We denote their
backgrounds as $\delta C_4=b_0^I\omega_4^I+\cdots$ with 
$I=1,\cdots h_{1,1}^+$.\footnote{$\cdots$ represent terms needed
for self-duality of $\tF_5$.}  4D vectors
can also descend from $C_4$ and are associated with positive parity harmonic
3-forms, which can appear in O7-plane compactifications.

Since we will consider only stabilized, constant $\tau$, 
we combine the 2-forms into a 
single complex 2-form $A_2$ with negative parity even at linear order in 
perturbations.  Backgrounds of the form
$\delta A_2=a_0^i \omega_2^i$ with $\omega_2^i$ a 
harmonic representative of $H_{1,1}^-$ 
yield axion-like moduli ($i=1,\cdots h_{1,1}^-$).
\comment{\footnote{As we will see later, though, when
$(\omega_2)^2$ is not harmonic, there must be an accompanying fluctuation in
$C_4$ even ignoring warping and flux.} }  These moduli are not stabilized by
the 3-form flux in general, and we present their dimensional reduction below
for the first time.  We might also consider scalars dual to 
$\delta A_{\mu\nu}$ in 4D,
but these components are removed by the orientifold projection.
In principle, 4D vectors can descend from $A_2$, as well, but they have no 
zero modes on a generic CY due to the trivial first cohomology.

The stabilized moduli (axio-dilaton and complex structure) have masses set
by the background flux. In the large-volume limit where the warp factor becomes
constant, this mass is parametrically lower than the Kaluza-Klein scale
(specifically, $m_{flux}^2/m_{KK}^2\sim e^{2\Omega}$).  This hierarchy implies any
symmetry or supersymmetry breaking by flux is spontaneous, which allows us to
use 4D supergravity to describe the effective theory.  In the supergravity
formalism, moduli 
stabilization by flux can be summarized by the superpotential 
$W\sim\int G_3\Omega_3$ \cite{hep-th/9906070,hep-th/0105097}.
We are concerned with finding the K\"ahler potential for
the unstabilized moduli in the 4D supergravity.

\subsection{Notes on torus factors}\label{torus}

As is well-known, generic CY manifolds have vanishing first Betti number and
no isometries.  However, since they are Ricci-flat, compactifications with
torus factors also solve the ansatz (\ref{gkpbackground}).  Specifically,
the three possibilities are $T^6/\mathbb{Z}_2$ O3 orientifolds as studied in
\cite{hep-th/0201028,hep-th/0201029} and O7 orientifolds 
$K3\times T^2/\mathbb{Z}_2$ and $T^4\times T^2/\mathbb{Z}_2$ \cite{hep-th/0301139}.
(We consider allowed orbifolds of these cases to be generic CY.) Due to
the fact that the tori do have harmonic 1-forms and isometries, 
there are a few additional features for compactifications with torus factors 
in comparison to the generic CY case.

First, isometries on a torus factor do allow zero modes corresponding to
Kaluza-Klein graviphotons.  However, they survive
the orientifold projection only on $T^4\times T^2/\mathbb{Z}_2$ compactifications.

The nonvanishing first Betti number has interesting consequences in the 
presence of flux.  Because there are nontrivial harmonic 5-forms on these 
compactifications, primitivity ($\t J\G=0$) is not guaranteed as it is on a 
generic CY.  As a result, starting in a supersymmetric compactification,
deformations of the K\"ahler structure by a harmonic
form $\omega_2$ with
$\omega_2\G\neq 0$ break supersymmetry and are no longer moduli (in supergravity
language, their mass is due to $D$-terms \cite{hep-th/0201028}).  
While we do
not consider K\"ahler moduli in this paper, we do consider their axionic
partners.  What happens to $\delta C_4=b_0(x)\t\star\omega_2+\cdots$ 
in this nonprimitive case? 
Because $\G$ is ISD, we have that 
$(\t\star\omega_2)_{mnpq}\G{}^{\widetilde{npq}}\neq 0$.  On torus-factor 
compactifications, it is straightforward to argue that $\t\star\omega_2$
can be written as $\lambda_1\G+\b\lambda_1\bG+\omega'_4$, where the contraction
of $\G$ with $\omega'_4$ vanishes.  We can therefore concern ourselves only with
moduli of the form $\delta C_4=b_0(x)(\lambda_1\G+\b\lambda_1\bG)+\cdots$.  From
the gauge transformation (\ref{gauge5}), however, we can see that such 
fluctuations in $C_4$ are pure gauge at linear order.  In fact, these are
Goldstone bosons for gauge fields $\delta A_2=a_1(x)\b\lambda_1(y)$, and
the $C_4$ fluctuations are eaten by the St\"uckelberg mechanism
\cite{hep-th/0201029,hep-th/0308156}.

We also note here an apparently hitherto unnoticed effect of flux on the
$A_2$ moduli, which occur only on the $T^4\times T^2/\mathbb{Z}_2$ 
compactifications among the types considered in this section.  
Consider a fluctuation associated with a negative parity 
harmonic 2-form $\omega_2$ with $\omega_2\G\neq 0$; the ``transgression'' 
(or Chern-Simons) terms
in $\tF_5$ as well as 5-form self-duality imply terms of the form
\beq{torusA2}
\delta\tF_5 \sim \frac{ig_s}{2} \left(a_0 \omega_2\w\bG -\b a_0\omega_2\w\G
\right)+\frac{ig_s}{2}e^{4\Omega_{\0}} e^{8A_{\0}} \h\epsilon\w
\left[a_0\t\star(\omega_2\w\bG)+\b a_0\t\star(\omega_2\w\G)\right]\ .\eeq
Even ignoring the warp factor, the second set of terms contributes to the
$G_3$ EOM as a mass term.  If $\omega_2\G$ were exact, these terms could be
removed by a shift of $\delta C_4$, but $\omega_2\G$ must be harmonic on a 
torus.  Alternately, the $|\tF_5|^2$ term in the action gives a mass term for
these modes.  Presumably, like K\"ahler moduli leading to non-primitive
flux, the masses for these moduli come from $D$-terms.
So we will not consider $A_2$ axions with 
harmonic (on the CY) $\delta A_2\G\neq 0$, since they are not moduli.

\section{Linearized EOM in warping}\label{warped}

In this section, we present the zero modes corresponding to each modulus 
considered: the universal volume modulus, the 4-form axions, and the 2-form
axions.  While we find that the zero mode for a single 4D degree of freedom
mixes several different 10D fields, we choose a gauge such that the static
perturbations take the form discussed in \S\ref{cyorientifolds} above.  
These are precisely the deformations that take one background into another
of the same class, and this gauge keeps $\t g_{mn}$ spacetime-independent and
Ricci-flat.  As mentioned in \S\ref{gkp}, there are generally 
background values for the $C_4$ and $A_2$ axion moduli, which do not appear
in field strengths but are not gauge trivial.

Before we proceed, we point out an important mathematical fact, which we
will emphasize in section \ref{formality} below.  In general, \textit{the wedge
product of harmonic forms need not be harmonic.}  As a result, the wedge product
of a harmonic 2-form with $\G$ may be a nonvanishing exact form on a generic
CY, a case we must consider carefully in our EOM.  We will see other consequences
of this fact, as well.

Since we need only the linearized EOM to find the kinetic action, we can study
each modulus separately.  We begin by presenting a review of the universal volume
modulus, followed by scalars descending from the 4-form potential and complex
2-form.

\subsection{Review of universal volume modulus}\label{volume}

Here we review briefly the dimensional reduction of the universal volume
modulus, as presented in \cite{arXiv:0810.5768}.  To first order, the metric and 
5-form can be written as
\beq{volumemetric}
ds^2 = e^{2\Omega} e^{2A} \h\eta_{\mu\nu} dx^\mu dx^\nu - 2e^{2\Omega}e^{2A}
\del_\mu c\del_m K(y)dx^\mu dy^m+e^{-2A}\tilde g_{mn}dy^m dy^n\eeq
and 
\beq{volumeF5}
\tF_5 = e^{4\Omega}\hat\epsilon\w\t de^{4A} +\t\star\t de^{-4A}-
e^{4\Omega}\hat\star\h d c\w \t dK\w \t de^{4A}\ ,\eeq
where the volume modulus $c(x)$ enters into the warp factor $A$ and the 
Weyl scaling factor $\Omega$, which should both be read as including
background and first-order parts.  The $x^\mu$ dependence of both fields
is solely implicit through this depencence on $c(x)$.
The 3-form is unmodified from the background.  

As these are special cases of 
the forms (\ref{generalBmetric},\ref{generalB5form}), we can use the
results of appendix \ref{offdiagonal} to write the constraints
from the $E_{\mu\nu}$ and $E_{mn}$ Einstein equations as
\bea
\del_\mu\del_\nu c\left\{ \t\Del^{\t 2}K -e^{-4A}\frac{\del e^{-2\Omega}}{\del c}
+e^{-2\Omega}\frac{\del e^{-4A}}{\del c}\right\}&=&0\nonumber\\
\del_\mu c \del_m \frac{\del e^{-4A}}{\del c} &=&0\ . \label{volumeconstraints}
\eea
The second of these enforces a form $e^{-4A}=e^{-4A_{\0}(y)}+f(c(x))$, which 
is consistent with the instantaneous solution of the background EOM for the
warp factor; for simplicity, we can choose $f(x)=c$.  Then we find that
$e^{-2\Omega(x)}=c(x)+e^{-2\Omega_{\0}}$ and 
\beq{volumeconstraint2} \t\Del^{\t 2}K(y) =e^{-4A_{\0}(y)} -e^{-2\Omega_{\0}}\ .\eeq
In this form, we can set the VEV of $c$ to zero; a nonzero VEV just changes the
values of $A_{\0}$ and $\Omega_{\0}$.
The constraints from the $\tF_5$ and $G_3$ EOM are automatically satisfied
(in particular, in the latter, the first-order contribution from the 5-form
cancels with a first-order contribution from the 10D Hodge star).

In this form, the first-order fields are
\bea
\delta g_{\mu\nu} &=& -e^{2\Omega_{\0}}e^{2A_{\0}} \left(e^{2\Omega_{\0}}+\frac 12 e^{4A_{\0}}
\right)c(x)\h\eta_{\mu\nu}\ ,\quad
\delta g^{\mu\nu}=\left(e^{-2A_{\0}}+\frac 12 e^{-2\Omega_{\0}}e^{2A_{\0}}
\right)c(x)\h\eta^{\mu\nu}\nonumber\\
\delta g_{\mu m} &=& -e^{2\Omega_{\0}}e^{2A_{\0}} \del_\mu c\,\del_m K\ ,\quad
\delta g^{\mu m} = e^{2A_{\0}} \del^{\h\mu} c\,\del^{\t m} K\nonumber\\
\delta g_{mn} &=& \frac 12 e^{2A_{\0}} c(x) \t g_{mn}\ ,\quad
\delta g^{mn}=-\frac 12 e^{6A_{\0}} c(x) \t g^{mn}\ .\label{volumeperturbedfields}
\eea
The 4-form potential can be written in a gauge such that we exclude all
the components in dimensional reduction following our prescription for
self-duality.  Similarly, the dynamical EOM as presented in appendix
\ref{IIBEOM} are
\bea
\delta E_{mn} &=& \del^{\h 2} c\left\{ \t\Del^{\t 2}K \t g_{mn}-\t\Del_m\del_n K
-4\del_{(m}A\del_{n)} K+2\del^{\t p}A\del_pK\t g_{mn}
-\frac 32 e^{-4A}\t g_{mn}+\frac 12 e^{-2\Omega}\t g_{mn}\right\}\nonumber\\
\delta E_6 &=& -e^{4\Omega}\h d\h\star\h d c\w \t d K\w \t de^{4A}\nonumber\\
\delta E_8 &=& -ie^{4A_{\0}}\h d\left(e^{4\Omega_{\0}}\h\star\h dc\right)\w\t dK\w\G 
+\frac i2 a_0^i\omega_2^i\w\delta E_6\ .
\label{volumeperturbedEOM}\eea
Note that, due to the overall first-order factor of $c(x)$ in both equations,
any appearance of $\Omega$ or $A$ in (\ref{volumeperturbedEOM}) takes
the background value.

As shown in \cite{arXiv:0810.5768}, the resulting kinetic term in 4D is 
\beq{volumeaction} S = -\frac{3\t V}{4\kappa^2} 
\int d^4x\, e^{4\Omega_{\0}} \del_\mu c\del^{\h\mu} c=-\frac{1}{\kappa_4^2}\frac{3}{4}
\int d^4x\, e^{4\Omega_{\0}} \del_\mu c\del^{\h\mu} c\ .\eeq
Up to a constant shift of the field $c$ by $e^{-2\Omega_{\0}}$, this is exactly the
same as the kinetic term for the volume modulus in an unwarped CY 
compactification, as is required by the no-scale structure of the background.
However, in cases with multiple moduli, no-scale structure is not so 
restrictive.

We close this review with a short comment.  It is possible to choose a 
gauge in which $g_{\mu m}=0$ and the compensator field $K(y)$ appears in
$g_{\mu\nu}$.  This gauge takes precedence over the ``off-diagonal metric gauge''
in the sense that it can be promoted to a full nonlinear solution of the
supergravity, at least for null waves in spacetime \cite{arXiv:0810.5768}.  
Nonetheless, we present here a form parallel to that required for axion
moduli.

\subsection{\boldmath Axions from $C_4$}\label{C4axions}

In this section, we discuss the constraints and EOM for 4D scalar degrees 
of freedom that descend from the 10D 4-form $C_4$.  Due to the self-duality
of $\tF_5$, these scalars appear in components of $C_4$ with all legs internal
or two external and two internal; we will describe these scalars as descending 
from components with all legs along the compact dimensions, as consistent with
5-form self-duality.  These scalars have only derivative couplings and comprise the
imaginary parts of the CY K\"ahler moduli, so they are axions.\footnote{A 
solution for the axionic partner of the universal volume modulus was
presented in \cite{arXiv:0810.5768}; we correct a sign error in that solution
and generalize our results to all the K\"ahler moduli axions.}
In terms of the 10D fields, a spacetime-independent shift 
$\delta C_4=b_0^I\omega_4^I$
with $\omega_4^I$ harmonic ($I=1,\cdots h_{1,1}^+$) leaves $\tF_5$ unchanged; we
now promote $b_0^I$ to spacetime fields.  As we discussed in section \ref{torus},
if $(\t\star\omega_4^I)\G\neq 0$ on a torus, such a mode is actually a 
Goldstone boson; we do not consider Goldstone bosons in this paper, concentrating
on moduli.

In this system, constraints arise from three sources.  First, since $C_4$ is
notrivial in the background, the $C_4$ gauge transformation leads to a constraint
among the 10D $\tF_5$ EOM.  As a result, a compensator field appears in the 
fluctuation of $C_4$, which we write as 
\beq{b0deltaC4}
\delta C_4 = b_0^I(x)\w\omega_4^I-\h d b_0^I \w K_3^I(y)+
\frac{ig_s}{4}\h d b_0^I\w \left(a_0^i\omega_2^i\w\Lambda_1^I-
\b a_0^i\omega_2^i\w \b\Lambda_1^I\right)+\cdots\ ,\eeq
where $\cdots$ represent terms needed to preserve self-duality of $\tF_5$ but which
we must discard in dimensional reduction in order to implement that self-duality.
The last term can be absorbed into the definition of $K_3^I$; we write 
it explicitly for later notational simplicity.
Note that a constant value of $b_0^I$ is allowed as a background value, but we
write it as part of the fluctuation since only its derivatives can appear in
gauge-invariant quantities.
Also, since $C_4$ transforms under the $A_2$ gauge transformation in the presence of
$G_3$, the $G_3$ EOM contain a constraint, and we introduce a compensator
\beq{b0deltaA2} \delta A_2 =-\h db_0^I\w \Lambda_1^I(y)\ .\eeq
We will see that this constraint and compensator become trivial in the absence of
a background $G_3$ flux.
Finally, 10D diffeomorphisms in the presence of a warp factor and background form
fields lead to a constraint in the Einstein equations, so we introduce a metric
compensator
\beq{b0deltametric} \delta g_{\mu m}=e^{2\Omega_{\0}}e^{2A_{\0}}\del_\mu b_0^I B_m^I(y)
\ ,\quad \delta g^{\mu m}=-e^{2A_{\0}}\del^{\h\mu} b_0^I B^{\t m,I}(y)
\ .\eeq
As denoted in the above, $K_3^I, \Lambda_1^I, B_1^I$ are forms on the internal CY.
For notational simplicity, we will henceforth suppress the index $I$
when it can be understood by context; furthermore,
the Weyl scaling factor and warp factor both take their background values, so we 
will omit the understood subscript $\0$ in this section.

Including the Chern-Simons terms, the 5-form flux is 
\bea
\tF_5 &=& e^{4\Omega}\h\epsilon\w\t de^{4A} +\t\star\t d e^{-4A} 
+\h d b_0\w\left( \omega_4+\t d K_3-\frac{i g_s}{2}
\left(\Lambda_1\w \b G_3^{\0}-\b\Lambda_1\w G_3^{\0}\right)\right)
\phantom{+e^{4\Omega}\h\star\h d b_0\w B_1\w d}\nonumber\\
&&+e^{4\Omega}\h\star\h d b_0\w B_1\w\t d e^{4A}
+e^{2\Omega}e^{4A}\h\star\h db_0\w\t\star\left( \omega_4+\t d K_3-\frac{i g_s}{2}
\left(\Lambda_1\w \b G_3^{\0}-\b\Lambda_1\w G_3^{\0}\right)\right)\ .\eea
The terms on the second line lead to both a dynamical EOM and a nontrivial
constraint
\beq{C4F5constraint1}
\t d\left[e^{4A}\t\star\left( \omega_4+\t d K_3-\frac{i g_s}{2}
\left(\Lambda_1\w \b G_3^{\0}-\b\Lambda_1\w G_3^{\0}\right)\right)+e^{2\Omega}
B_1\w\t de^{4A}\right]=0\ ,\eeq
which is solved for
\beq{C4F5gammadef1}
e^{4A}\left[\t\star\left( 
\omega_4+\t d K_3-\frac{i g_s}{2}
\left(\Lambda_1\w \b G_3^{\0}-\b\Lambda_1\w G_3^{\0}\right)\right)+
e^{2\Omega}\t dB_1\right]=\gamma_2\eeq
with $\gamma_2$ closed.
The integrability condition for this identification is therefore
\beq{C4F5Bpoisson}
e^{2\Omega}\t d\t\star \t d B_1 = \t d e^{-4A}\w \t\star \gamma_2+e^{-4A}
\t d\t\star \gamma_2+
\frac{ig_s}{2} \left(\t d\Lambda_1\w\b G_3^{\0}-\t d\b\Lambda_1\w G_3^{\0}
\right)\ ,\eeq
which is equivalent to the constraint (\ref{C4F5constraint1}).
We will return to the question of how equation (\ref{C4F5gammadef1}) 
determines $\gamma_2$ below.  The dynamical equation of motion
can be written as
\beq{C4F5perturbedEOM}
\delta E_6 = %\h d\left(e^{2\Omega}\h\star\h db_0\right)\w \left[ e^{4A}\t\star
%\left(\omega_4+\t dK_3-\frac{ig_s}{2}\Lambda_1\w \b G_3^{\0}+
%\frac{ig_s}{2}\b\Lambda_1\w G_3^{\0}\right)+e^{2\Omega}B_1\w\t de^{4A}\right]
%\nonumber\\
\h d\left(e^{2\Omega}\h\star\h db_0\right)\w \left[\gamma_2-
e^{2\Omega}\t d\left(e^{4A}B_1\right)\right]\ .\eeq

Our fields follow the ansatz given in appendix \ref{offdiagonal}, so
the Einstein equations as derived there yield the constraints
$\t\Del^{\t m}B_m=0$ 
($\mu\nu$ component), which implies that $B_1$ is co-exact, and
(from the $\mu m$ component)
\bea
\t\Del^2 B_m -4(\t d B)_{mn}\del^{\t n}A&=& 4e^{-2\Omega}\t\star
\left( \omega_4+\t d K_3-\frac{i g_s}{2}
\left(\Lambda_1\w \b G_3^{\0}-\b\Lambda_1\w G_3^{\0}\right)\right)_{mn}\del^{\t n}A
\nonumber\\ 
&&-\frac{i g_s}{2} \t\star\left(\t d\Lambda_1\w\b G_3^{\0}-\t d\b\Lambda_1
\w G_3^{\0}\right)_m\ \ \textnormal{or}\nonumber\\
\t\Del^2 B_m &=& 4e^{-2\Omega}e^{-4A}\gamma_{mn}\del^{\t n}A 
-\frac{i g_s}{2} \t\star\left(\t d\Lambda_1\w\b G_3^{\0}-\t d\b\Lambda_1
\w G_3^{\0}\right)_m\, .\label{C4einsteinpoisson}
\eea
This equation is consistent with (\ref{C4F5Bpoisson}) when $\gamma_2$ is
harmonic.  With these constraints satisfied, the dynamical EOM are
all in the $mn$ component, which is
\beq{C4einsteinEOM}
\delta E_{mn} = \del^{\h 2} b_0 \left(\t\Del_{(m}B_{n)}+4\del_{(m}AB_{n)} -
2\t g_{mn} B_p\del^{\t p} A\right)\ .
\eeq

We finally must consider the $G_3$ EOM $d\star G_3 = -iG_3 \tF_5$.  
After some simplification, the constraint becomes
\beq{C43formpoisson}
-\t d\t\star \t d\Lambda_1 =i\gamma_2\w G_3^{\0}\quad \textnormal{or}\quad 
\t\Del^2\Lambda_m =-\frac 1 2 G^{\0}_{mnp}\gamma^{\widetilde{np}}
\eeq
in a gauge with $\t d\t\star\Lambda_1=0$.  This equation implies that 
$\gamma_2 G_3^{\0}$ is purely exact, which is automatically true on a CY 3-fold
because the fifth Betti number vanishes.  On a $T^6/\mathbb{Z}_2$ or
$K3\times T^2/\mathbb{Z}_2$ orientifold, a non-vanishing harmonic part makes this
equation insoluble; we argue in \S\ref{formality} that this occurs
exactly when the corresponding
axion is in fact a Goldstone mode for a massive vector descending from $A_2$,
which would therefore require a different solution to the 10D EOM.  We
henceforth assume $\gamma_2 G_3^{\0}$ is trivial in cohomology.  Note, though,
that $\Lambda_1$ may be nontrivial even in the large-volume limit when the 
warp factor approaches a constant on the CY.
The dynamical EOM associated with $A_2$ is
\beq{C43formEOM}\delta E_8 =\h d\left(e^{2\Omega}\h\star\h d b_0\right)\w
\left(\t\star \t d \Lambda_1 +ie^{2\Omega} e^{4A}B_1\w G_3^{\0}\right)
+\frac i2 a_0^i\w\omega_2^i\w\delta E_6\ .\eeq

The remaining issue is determining the values of $B_1$, $\Lambda_1$, and $\gamma_2$
given the background geometry and a choice of $\omega_2$.  
$\Lambda_1$ is determined by the Poisson-like equation (\ref{C43formpoisson})
in terms of $\gamma_2$ and the background, while $B_1$ is similarly
determined by a Poisson-like equation (\ref{C4einsteinpoisson}) in terms
of $\Lambda_1$, $\gamma_2$, and background fields.  Likewise, 
(\ref{C4F5gammadef1}) defines $K_3$ (and gives a Poisson-like equation for it).  
We assume that these equations are
soluble since they all have Poisson-like form, 
so what remains is finding $\gamma_2$ from $\omega_4$.  Since
this process is similar for $A_2$ axions, we discuss it and
related mathematical issues below in section \ref{formality}.

\subsection{\boldmath Axions from $A_2$}\label{A2axions}

Here we consider the zero modes of scalars that descend from the 10D 2-form
potential $A_2$, which are associated with harmonic forms $\omega_2^i$ with
odd intrinsic parity under the orientifold involution ($i=1,\cdots h_{1,1}^-$).  
Due to the transgression
terms in $\tF_5$, these scalars can have non-derivative couplings.  Specifically,
unlike axions descending from $C_4$, the VEV of these scalars appears in kinetic
terms; it is possible to make a shift symmetry in either the real or
imaginary part of $a_0^i$ manifest with a field redefinition, but not both.  
Nonetheless, these moduli are commonly referred to as axions.  As noted earlier,
modes with harmonic $\omega_2^i \G\neq 0$ gain mass; this case only occurs on 
toroidal compactifications, and we do not consider it further.  On general
CY manifolds, however, we do allow $\omega_2^i \G\neq 0$ exact and see that the
corresponding modes are classically massless.  Nonperturbative stabilization of
these axions has recently been considered in \cite{Gao:2013pra,Gao:2013rra}.

As it turns out, the 10D EOM of these is similar to that for axions that 
descend from $C_4$, the main difference occuring when 
$\omega_2^i\G=\t dQ_4^i\neq 0$: 
constraints arise from gauge transformations of $A_2$ and
$C_4$ as well as diffeomorphisms, and there are nontrivial compensators as in
\S\ref{C4axions}.  Omitting the terms required to generate the flux $G_3^{\0}$, 
we take a background $A_2^{\0}=a_0^i\omega_2^i$ with
\beq{a0deltaA2} \delta A_2 =\delta a_0^i(x)\w\omega_2^i -\alpha_1^{ij} \w
\Lambda_1^{ij}\ ,\quad \alpha_1^{ij}\equiv\frac{ig_s}{4}\left(a_0^i\h d\delta\b a_0^j
-\b a_0^i\h d\delta a_0^j\right)\ .\eeq
With $a_0^i(x)=a_0^i+\delta a_0^i(x)$ as the 4D field, $\h da_0^i=\h d\delta a_0^i$,
so we will omit the leading $\delta$ on fluctuations when differentiated.
Note also that, to first order, $\h d\alpha_1^{ij}=0$.

We noted in section \ref{torus} that $\omega_2^i\G\neq 0$ can lead to a mass
term; however, this term can be removed from $\tF_5$
for $\omega_2^i\G=\t dQ_4^i$ by an 
appropriate shift of $C_4$ by $-(ig_s/2)a_0^i\b Q_4^i+cc$.  Therefore,
we take
\bea
\delta C_4 &=& -\alpha_1^{ij} \w K_3^{ij} +\frac{ig_s}{4}
\left( a_0^i\w \alpha_1^{jk}\w\omega_2^i\w\Lambda_1^{jk}-
\b a_0^i\w \alpha_1^{jk}\w\omega_2^i\w\b\Lambda_1^{jk}\right)\nonumber\\
&&-\frac{ig_s}{2} \left(\delta a_0^i\w\b Q_4^i-\delta \b a_0^i\w Q_4^i\right)
+\cdots\label{a0deltaC4} 
\eea
along with a metric compensator
\bea
\delta g_{\mu m}&=&e^{2\Omega_{\0}}e^{2A_{\0}}\left(\alpha_\mu^{ij} B_m^{ij}(y)
-\frac{ig_s}{2} \del_\mu a_0^i\beta_m^i(y)+\frac{ig_s}{2}\del_\mu\b a_0^i
\b\beta_m^i(y)\right)
\ ,\nonumber\\ 
\delta g^{\mu m}&=&-e^{2A_{\0}}\left(\alpha^{\h\mu,ij} B^{\t m,ij}
-\frac{ig_s}{2} \del^{\h\mu} a_0^i\beta^{\t m,i}+\frac{ig_s}{2}
\del^{\h \mu}\b a_0^i\b\beta^{\t m,i}\right)
\ .\label{a0deltametric} \eea
Since $A$ and $\Omega$ are not modified for this axion motion, we omit
the subscript $\0$ for the background values again in this section.

With this form for $\delta C_4$, $\tF_5$ becomes
\bea
\tF_5&=&e^{4\Omega}\h\epsilon\w\t de^{4A} +\t\star\t d e^{-4A} 
+\alpha_1^{ij}\w\left( \omega_2^i\w\omega_2^j+\t d K_3^{ij}-\frac{i g_s}{2}
\left(\Lambda_1^{ij}\w \b G_3^{\0}-\b\Lambda_1^{ij}\w G_3^{\0}\right)\right)
\phantom{+e^{4\Omega}\h\star\h d b_0\w B_1\w d}\nonumber\\
&&+e^{2\Omega}e^{4A}\h\star\alpha_1^{ij}\w\left[
\t\star\left( \omega_2^{i}\w\omega_2^{j}+\t d K_3^{ij}-\frac{i g_s}{2}
\left(\Lambda_1^{ij}\w \b G_3^{\0}-\b\Lambda_1^{ij}\w G_3^{\0}\right)\right)
+e^{2\Omega}B_1^{ij}\w\t de^{4A}\right]\nonumber\\
&&-\frac{ig_s}{2} \left(\h da_0^i \w\b Q_4^i -\h d\b a_0^i \w Q_4^i\right)
-\frac{ig_s}{2} e^{2\Omega}\left[\h\star\h d a_0^i \left( e^{4A}\t\star\b Q_4^i
+e^{2\Omega} \beta_1^i\w\t d e^{4A}\right)\right.\nonumber\\
&&\left.-\h\star\h d\b a_0^i \left( e^{4A}\t\star Q_4^i
+e^{2\Omega} \b\beta_1^i\w\t d e^{4A}\right) \right]\ .
\label{a0F5}
\eea
The first two lines are clearly very similar to the compensators required 
for the $C_4$ 
axions, so we can summarize the constraints briefly as follows.  Suppressing
$i,j$ indices, the $\tF_5$ constraint requires
\beq{A2gammadef}
e^{4A}\left[\t\star\left( 
\omega_2\w\omega_2+\t d K_3-\frac{i g_s}{2}
\left(\Lambda_1\w \b G_3^{\0}-\b\Lambda_1\w G_3^{\0}\right)\right)+
e^{2\Omega}\t dB_1\right]=\gamma_2\eeq
for $\gamma_2$ harmonic.  The integrability condition for this equation and
the Einstein equation yield the same Poisson equation
\beq{A2poisson}
\t\Del^2 B_m = 4e^{-2\Omega}e^{-4A}\gamma_{mn}\del^{\t n}A 
-\frac{i g_s}{2} \t\star\left(\t d\Lambda_1\w\b G_3^{\0}-\t d\b\Lambda_1
\w G_3^{\0}\right)_m\, .\eeq
Like $B_1^I$, $B_1^{ij}$ is co-exact and divergenceless.
There is an additional constraint from the $\h\star\h d a_0$ terms, which can
be written as 
\beq{betaQ}
\t d\left[ e^{4A}\left( \t\star\b Q_4+e^{2\Omega} \t d\beta_1\right)-e^{2\Omega}
\t d\left(e^{4A}\beta_1\right)\right] =0\ . 
\eeq
There is an interesting solution to this constraint; $Q_4$ must have a co-exact part
and can be written as $Q_4\t\star\t d\t\star\theta_5$.  Then 
$\beta_1 =-e^{-2\Omega}\t\star\b\theta_5$ solves the constraint!
The final constraint from the $G_3$ EOM is identical to equation 
(\ref{C43formpoisson}) --- $\beta_1$ and $Q_4$ do not enter.  (As a note,
we can formally solve (\ref{C43formpoisson}) using $\theta_5$ as here.)

Finally, it is interesting to note that $K_3^{ij}\neq 0$ even in the absence of
warping and flux when all other compensators vanish, as $\omega_2^i\omega_2^j$
need not be the harmonic representative of its cohomology class.

The dynamical EOM for these moduli are
\bea
\delta E_{mn} &=& \del^{\h\mu} \alpha_\mu^{ij}\left(\t\Del_{(m}B_{n)}^{ij}+4
\del_{(m}AB_{n)}^{ij} -2\t g_{mn} B_p^{ij}\del^{\t p} A\right)
-\frac{ig_s}{2} \del^{\h 2}a_0^i\left(\t\Del_{(m}\beta_{n)}^i+4\del_{(m}A\beta_{n)}^i
-2\t g_{mn}\beta_p^i\del^{\t p} A\right) \nonumber\\
&&+\frac{ig_s}{2} \del^{\h 2}\b a_0^i\left(\t\Del_{(m}\b \beta_{n)}^i
+4\del_{(m}A\b \beta_{n)}^i -2\t g_{mn}\b\beta_p^i\del^{\t p} A\right)\nonumber\\
\delta E_6&=&\h d\left(e^{2\Omega}\h\star\alpha_1^{ij}\right)\w\left[
\gamma_2^{ij}-e^{2\Omega}\t d\left(e^{4A}B_1^{ij}\right)\right]
+\frac{ig_s}{2}\h d\left(e^{4\Omega}\h\star\h d a_0^i\right)\w
\t d\left(e^{4A}\beta_1^i\right)\nonumber\\
&&-\frac{ig_s}{2}\h d\left(e^{4\Omega}
\h\star\h d\b a_0^i\right)\w\t d\left(e^{4A}\b\beta_1^i\right)\nonumber\\
\delta E_8 &=& \hat d\left(e^{2\Omega}\h\star\h da_0^i\right)\w\t\star\omega_2^i
+\h d\left(e^{2\Omega}\h\star\alpha_1^{ij}\right)\w\left[\t\star\t d\Lambda_1^{ij}
+ie^{2\Omega}e^{4A}B_1^{ij}\w
\G\right]\nonumber\\
&&+\frac{g_s}{2}\h d\left(e^{4\Omega}\h\star\h d a_0^i\right)\w\left[e^{4A}
\beta_1^i\w\G\right]
-\frac{g_s}{2}\h d\left(e^{4\Omega}\h\star\h d\b a_0^i\right)\w\left[e^{4A}
\b\beta_1^i\w\G\right]
+\frac i2 a_0^i\w\omega_2^i\w\delta E_6\, .\label{A2dynamical}
\eea

\subsection{\boldmath Properties of harmonic forms and $\gamma_2$}\label{formality}

We have left unanswered the question of how to determine $\gamma_2^I$ in terms
of $\omega_4^I$ for the $C_4$ axions and $\gamma_2^{ij}$ in terms of 
the $\omega_2^i$ for the $A_2$ axions.  
\comment{In fact, it is more straightforward to define the 
axion modulus in terms of
the associated form $\gamma_2$ and determine $\omega_{2,4}$ in terms of $\gamma_2$.}
Here we explain the process and related mathematical questions.

Both axions are very similar, so we present the $C_4$ axion case in 
detail and summarize $A_2$ axions later.
Choose $\omega_4^I=\t\star\omega_2^I$ and 
$\gamma_2^I=C^{IJ}\omega_2^J$ in terms of %to be %proportional to 
the basis $\{\omega_2^I\}$ of positive orientifold parity harmonic $(1,1)$
forms, %and write the associated 4-form as $\omega_4^I=C^{IJ}\t\star\gamma_2^J$.
and note that $\Lambda_1^I=C^{IJ}\check\Lambda_1^J$ where $\check\Lambda_1^J$
satisfies (\ref{C43formpoisson}) for $\gamma_2=\omega_2^J$.
\comment{For the usual field definition where $\omega_4^I$ is the star of a basis 
(1,1) form, we take appropriate linear combinations.}
If we wedge equation (\ref{C4F5gammadef1}) with $\t\star\omega_2^J$, we find
\beq{C4F5gammadef2}
3\t V \delta^{IJ} = C^{IK}\left[\int e^{-4A_{\0}} \omega_2^K\w\t\star\omega_2^J
+\frac{ig_s}{2}\int\left(\check\Lambda_1^K\w\b G_3^{\0}-
\b{\check\Lambda}_1^K\w G_3^{\0}\right)\w \omega_2^J\right]\ .\eeq
If we define, as above, $\omega_2^I\G=\t dQ_4^I$, we can remove the explicit
compensators by rewriting (\ref{C43formpoisson}) as 
$\t\star\t d\check\Lambda_1=-iQ_4+\t d\lambda_3$:
\beq{C4F5gammadef3}
3\t V \delta^{IJ} = C^{IK}\left[\int e^{-4A_{\0}} \omega_2^K\w\t\star\omega_2^J
-\frac{g_s}{2}\int\left(Q_4^K\w\t\star\b Q_4^J+\b Q_4^K\w\t\star Q_4^J
\right)\right]\ ,\eeq
which is manifestly symmetric.
$C^{IJ}$ is then found by inverting the matrix in $KJ$ given by the integrals
in square brackets.
Since $\omega_4^I$ has positive orientifold parity, we can see that $\gamma_2^I$
must also; the negative parity equivalent of (\ref{C4F5gammadef3}) would require the
first integral in square brackets, which includes the warp factor, 
to cancel against the second, which does not.  Since the warp factor depends on the 
expectation value of the universal volume modulus while $Q_4^I$ does not,
there can be no part of $\gamma_2^I$ with negative parity.  This argument
uses the fact that the contraction of a negative parity form with a positive
parity form vanishes.

To isolate the effect of the warp factor, consider a case in which 
$\gamma_2 G_3^{\0}=0$, so $\Lambda_1=0$ (alternately, we can set $G_3^{\0}=0$).
Nonvanishing $\Lambda_1^I$ will provide an additional correction to this
discussion.
The key question is now evaluating $\int e^{-4A_{\0}}\omega_2^I\t\star\omega_2^J$.
This integral is related to the pointwise inner product of harmonic 2-forms,
$f^{IJ}(y) =\t\star(\omega_2^I\t\star\omega_2^J)$.  If $f^{IJ}(y)$ is 
constant, then it factors out of the integral, leaving
\beq{pointwiseinnerconst} \int e^{-4A_{\0}}\omega_2^I\w\t\star\omega_2^J \propto
3\delta^{IJ}\int e^{-4A_{\0}}\t\epsilon =3\t V e^{-2\Omega_{\0}}\delta^{IJ}\eeq
(since our normalization requires $f^{IJ}=3\delta^{IJ}$).  The conditions
under which the pointwise inner product is constant has been studied in a
small mathematical literature (see for example \cite{nagy2004,nagy2006,grosjean}), 
and generally it is not
necessarily constant.  While we have not found a definite statement regarding
harmonic forms on CY 3-folds, the pointwise inner product is certainly 
\textit{not} constant for harmonic forms on K3, since some harmonic forms 
vanish at special points \cite{kotschick2000}.  Indeed, if the 2nd Betti 
number is greater than the second Betti number on $T^6$, the pointwise inner
products of harmonic 2-forms cannot all be constant \cite{kotschick2000}.
Because the pointwise inner product is \textit{not}
generally constant, we therefore expect on CY with sufficiently large $h_{1,1}$
that \textit{a nontrivial warp factor
modifies the relation between $\gamma_2^I$ and $\omega_2^I$.}

In turn, the question of when the pointwise inner produce of harmonic forms 
is necessarily constant is related to the question of when the wedge product
of harmonic forms is necessarily harmonic, which need not hold because 
the co-derivative is not a derivation.  This property of a metric is
known as (metric) \textit{formality}, 
and harmonic forms always have constant pointwise
inner products in a formal metric.  Manifolds which support a formal metric
are \textit{geometrically formal}; the reader should not confuse geometric
formality with topological (aka Sullivan) formality, which is a necessary but
not sufficient condition for geometric formality and which all K\"ahler
manifolds possess \cite{deligne} (see \cite{kotschick2003,kotschick2011} for
some examples of topologically but not geometrically formal manifolds).  
As noted, the Ricci-flat metric on K3 is not formal, but
flat torus metrics are; generally, a manifold cannot be geometrically formal
if any of the Betti numbers are larger than the corresponding Betti numbers
on a torus of the same dimension \cite{kotschick2000}.  Further restrictions
on formal K\"ahler manifolds are found in \cite{nagy2006}.
  
In addition to warp factor effects, metric formality or the lack of it are
important to corrections from the flux.  A CY with the Ricci-flat
metric is not typically formal (and is often not geometrically formal), so 
$\gamma_2\G$ need not be harmonic.  In combination with the vanishing first
Betti number, that fact means that $\gamma_2\G$ can be a nonzero exact 5-form.
This has an interesting consequence for equation (\ref{C43formpoisson}); 
it allows a nonvanishing $\Lambda_1$, which affects the identification of 
$\gamma_2^I$.  %It is very important to note that equation (\ref{C43formpoisson})
%is independent of the volume modulus, so these flux-induced corrections
%do not vanish in the large-volume limit.  

We can, however, find solve $\gamma_2^{I=1}$ for $\omega_4^1=\t\star\t J$ given by 
the K\"ahler form of the CY, which is the universal axion, the partner of the
volume modulus (or the only axion if $h_{1,1}^+=1$).  
The point is that the K\"ahler form (the first element
of our basis of harmonic 2-forms) has a 
property much like formality: the wedge product of $\t J$ with any harmonic
form is itself harmonic (see the review by \cite{ballmann}).  
On a CY 3-fold, this is enough to imply that 
$\t J G_3^{\0}=0$, and the supersymmetry condition of primitivity does so
in some other cases.
Therefore, we easily find $C^{1J}=e^{2\Omega_{\0}}\delta^{1J}$, which
tells us that $\gamma_2^1=e^{2\Omega_{\0}}\t J$.  
This is the correct normalization to define $b_0$ as the 
imaginary part of a holomorphic coordinate on moduli space \cite{arXiv:0810.5768}.  

We can also consider the large-volume limit.  If we define 
$e^{-4A_{\0}(y)}=e^{-2\Omega_{\0}}+Z(y)$ and allow $e^{-2\Omega_{\0}}\to\infty$, the leading
term is $C^{IJ}=e^{2\Omega_{\0}}\delta^{IJ}$ (as for the CY without warping or flux),
and the corrections are
\beq{CIJlargevolume} \delta C^{IJ}=\frac{e^{4\Omega_{\0}}}{3\t V} \left[
\frac{g_s}{2} \int\left(Q_4^I\w\t\star\b Q_4^J+\b Q_4^I\w\t\star Q_4^J\right)-
\int\, Z(y)\omega_2^I\t\star\omega_2^J\right]\ ,\eeq
suppressed by a factor of $e^{2\Omega_{\0}}$.  Note, however, that the scale at which
$\delta C^{IJ}$ becomes small is not set by features of the CY manifold,
such as the volume of a 4-cycle, but rather by flux quanta and length of warped 
throat regions (themselves set by a combination of geometry and flux).

As mentioned, the two-form axion case is similar, except $\omega_2^i\in H_{1,1}^-$
while $\gamma_2^{ij}\in H_{1,1}^+$.  We again expand $\gamma_2^{ij}=c^{ijI}\omega_2^I$
and find 
\beq{A2gammadef2}
\int \omega_2^I\w\omega_2^i\w\omega_2^j 
= C^{ijJ}\left[\int e^{-4A_{\0}} \omega_2^J\w\t\star\omega_2^I
-\frac{g_s}{2}\int\left(Q_4^J\w\t\star\b Q_4^I+\b Q_4^J\w\t\star Q_4^I
\right)\right]\ .\eeq
With our normalization, the left-hand side of this equation is given
by the triple-intersection $3\t Vd^{Iij}$.  We again find significant simplification
in the case that $h_{1,1}^+=1$, regardless of the value of $h_{1,1}^-$.  Again,
$\check\Lambda_1=0$, and we can further write 
$\omega_2^i\omega_2^j=\beta^{ij}\t\star\t J$ in cohomology, so $d^{Iij}=\beta^{ij}$.
However, since orientifold involution parity implies that 
$\t J_{mn}\omega^{\widetilde{mn}}=0$, we can see that $\t\star\omega_2=-\t J\omega_2$.
With our normalization, therefore, $\beta^{ij}=-\delta^{ij}$.  Then 
$\gamma_2^{ij}=-e^{2\Omega_{\0}}\delta^{ij}\t J$.  In the more general case that
$h_{1,1}^+>1$, the common factor in (\ref{C4F5gammadef3},\ref{A2gammadef2}) implies
that $C^{ijI}=3\t V C^{IJ}d^{Jij}$.

On the other hand, 
for $T^6/\mathbb{Z}_2$ or $T^4\times T^2/\mathbb{Z}_2$ compactifications, 
$\gamma_2G_3^{\0}$ must be 
harmonic and therefore must vanish for equation (\ref{C43formpoisson}) to be
consistent.  However, using logic similar to that used to argue that
$\gamma_2$ has positive parity under the orientifold involution, we can see
that $\gamma_2\G\neq 0$ harmonic only when $\omega_2\G\neq 0$ harmonic.  
In that case, the putative axion is instead either massive or
a Goldstone boson. 
Formality of the torus metric and constancy of form inner products have
the additional
consequence that $\gamma_2^I\propto\omega_2^I$ on a torus orientifold 
compactification.

The key result of this section is that generally the presence of a warp factor
and 3-form flux modifies the relationship between 
$\gamma_2^I$ and $\omega_2^I$.  We will see what this means for the K\"ahler
metric and potential on moduli space in the following section.

\section{Moduli action and K\"ahler potential}\label{kahlerpotential}

To find the kinetic action on the moduli and therefore the metric on moduli
space, we follow the strategy laid out in section \ref{dimred} and appendix
\ref{linearizedeom}.  In order to find the complete kinetic action including
kinetic terms mixing degrees of freedom, we need to consider the complete
fluctuations of all the 10D fields superposing the three moduli described
above, as well as the superposed equations of motion.  As described in equation
(\ref{SUGRA2}), the quadratic action on moduli space is 
\bea
S &=& \frac{1}{4\kappa^2} \int d^{4}x\, e^{4\Omega_{\0}}\, \int d^6y\, \sqrt{\t g}\, 
e^{-2A_{\0}}\, \delta g^{MN}
\delta E_{MN}\nonumber\\ 
&&+\frac{1}{4\kappa^2}\int_{\mathbb{R}^{3,1}}\,\int_{CY}
\left(\delta C_4\w\delta E_6+
\frac{g_s}{2}\delta A_2\w\delta\b E_8 +\frac{g_s}{2}\delta\b A_2 \w\delta E_8 
\right)\ ,\label{kineticaction1}\eea
where $\kappa$ is the 10D Planck constant.  We will first carry out the integrals
over the CY manifold for our zero modes to convert this to a 4D action; then we
will discuss the K\"ahler potential that leads to the corresponding metric on
moduli space.

\subsection{Integrating out the Calabi-Yau}\label{intout}

Integrating (\ref{kineticaction1}) over the internal manifold allows us to
write an action for the 4D field theory that descends from the compactification,
as long as we enforce the constraints that arise from 10D gauge invariance
before integrating over the CY.  In addition, since massive modes are presumably
orthogonal to the zero modes, this integration allows us to consider only light
fields in our 4D theory.  We do not prove that the zero modes and massive
modes are orthogonal in this paper, but the subject was discussed in 
\cite{arXiv:0810.5768} for the volume modulus and universal axion.  
We are also ignoring any corrections due to interactions with massive modes,
including at tree-level (as in \cite{arXiv:1207.0815} for the potential in a 5D 
warped compactification) or threshold corrections.
This is the usual approximation used in discussions of warped string 
compactifications.

The integration is fairly straightforward.  For example, the 
Einstein equation term given by the first line of (\ref{kineticaction1})
receives nonzero contributions only from $\delta g^{mn}\delta E_{mn}$, which 
are (after contraction)
\bea
S_{Einstein} &=& -\frac{1}{8\kappa^2}\int d^4x\, e^{4\Omega_{\0}} c(x)
\int d^6y\,\sqrt{\t g}\, e^{4A_{\0}} \left\{\del^{\h 2}c(x) \left[5\t\Del^2 K
+8\del^{\t p}A_{\0}\del_p K -9 e^{-4A_{\0}}\right.\right.\nonumber\\
&&\left.\left.  +3e^{-2\Omega_{\0}}\right]+\del^{\h 2}b^I(x) 
\left[\t\Del^{\t p}B_p^I -8\del^{\t p}A_{\0} B_p^I\right]+
\del^{\h\mu}\alpha_\mu^{ij}\left[\t\Del^{\t p}B_p^{ij} -8\del^{\t p}A_{\0} B_p^{ij}
\right]\right\}\, .\label{kineticeinstein1}
\eea
Here and in the following, we drop the subscript 0 on the axions, denoting them
as any 4D scalar.  Since $B_1^I,B_1^{ij}$ are co-exact, it is simple to see
that the terms involving $b^I$ and $\alpha^{ij}$ vanish.  Integrating by parts
and imposing the constraints (\ref{volumeconstraints}), 
\beq{kineticeinstein2}
S_{Einstein} = 
-\frac{3\t V}{4\kappa^2}\int d^4x\, e^{4\Omega_{\0}}\del_\mu c
\del^{\h\mu} c\ ,\eeq
as claimed in section \ref{volume}.  Note that the factor of 3 appears precisely
due to application of the constraint.

The contribution from the form EOM are more interesting since they contribute
off-diagonal terms to the moduli space metric.  We note
that only components $\delta C_4$ and $\delta A_2$ with all internal legs
can contribute, since the dynamical EOM $\delta E_6$ and $\delta E_8$ both
have 4 external legs.  Therefore, we can write
\bea
\int \delta C_4\w\delta E_6 &=&\int \left[\delta b^I\omega_4^I-\frac{ig_s}{2}\left(
\delta a^i\b Q_4^i-\delta \b a^i Q_4^i\right)\right]\w\left[ -e^{4\Omega_{\0}}
\h d\h\star\h d c\w\t dK\w \t de^{4A_{\0}}\vphantom{\frac{ig_s}{2}}\right.\nonumber\\
&&\left. +e^{2\Omega_{\0}}\h d\h\star\h d b^J
\w\left(\gamma_2^J-e^{2\Omega_{\0}}\t d(e^{4A_{\0}}B_1^J)\right)
+e^{2\Omega_{\0}}\h d\h\star\alpha_1^{ij}\w\left(\gamma_2^{ij}
-e^{2\Omega_{\0}}\t d(e^{4A_{\0}}B_1^{ij})\right)\right.\nonumber\\
&&\left. +\frac{ig_s}{2} e^{4\Omega_{\0}}\h d\h\star\h d a^j\w\t d\left(e^{4A_{\0}}
\beta^j\right)-\frac{ig_s}{2} e^{4\Omega_{\0}}\h d\h\star\h d\b a^j\w
\t d\left(e^{4A_{\0}}\b\beta^j\right)\right]\nonumber\\
&=& -\int_{\mathbb{R}^{3,1}} e^{2\Omega_{\0}}\h d b^I\w\h\star\h db^J
\int_{CY}\omega_4^I\w\gamma_2^J -\int_{\mathbb{R}^{3,1}} e^{2\Omega_{\0}}\h d b^I\w
\h\star\alpha_1^{ij}\int_{CY}\omega_4^I\w\gamma_2^{ij}\nonumber\\
&&+\frac{ig_s}{2}\int e^{4\Omega_{\0}}\left[\h d a^i\w\b Q_4^i -\h d\b a^i\w Q_4^i
\right]\w\left[\h\star\h d c\w \t d\left(e^{4A_{\0}}\t d K\right)
-\h\star\h d b^I\w\t d\left(e^{4A_{\0}}B_1^I\right)\right.\nonumber\\
&&\left. -\h\star\alpha_1^{jk}\w\t d\left(e^{4A_{\0}}B_1^{jk}\right)
+\frac{ig_s}{2}\h\star\h d a^j\w\t d\left(e^{4A_{\0}}\beta_1^i\right)
-\frac{ig_s}{2}\h\star\h d\b a^j\w\t d\left(e^{4A_{\0}}\b\beta_1^i\right)
\right]\ .\label{kinetic4form1}
\eea
Many of the terms in the first line disappear upon integration by parts since
$\omega_4^I$ is harmonic and because $Q_4^i$ is co-exact.  
The calculation is similar for the $A_2$ EOM with
terms such as $\int \omega_2^i B_1\G$ cancelling with the $Q_4^i$ terms in 
(\ref{kinetic4form1}) after integration by parts.

After a little straightforward algebra, we find the kinetic action
\bea
S&=& -\frac{1}{\kappa^2} \int d^4x\left\{ \frac{3\t V}{4} e^{4\Omega_{\0}}
\del_\mu c\del^{\h\mu} c+ \frac 14 e^{2\Omega_{\0}}\del_\mu b^I\del^{\h\mu}b^J
\left(\int_{CY}\omega_4^I\w\gamma_2^J \right)\right.\nonumber\\
&&\left. +\frac{g_s}{4} e^{2\Omega_{\0}}
\del_\mu a^i \del^{\h\mu}\b a^j \left(\int_{CY}\omega_2^i\w\t\star\omega_2^j\right)
+\frac 14 e^{2\Omega_{\0}}\alpha_\mu^{ij}\alpha^{\h\mu\, kl}\left(\int_{CY}\omega_2^i
\w\omega_2^j\w\gamma_2^{kl}\right)\right.\nonumber\\
&&\left. +\frac 14 e^{2\Omega_{\0}}\del_\mu b^I\alpha^{\h\mu\, ij}\left(\int_{CY}
\left(\omega_4^I\w\gamma_2^{ij}+\omega_2^i\w\omega_2^j\w\gamma_2^I\right)\right)
\right\}\ .
\label{kineticaction2}\eea
Replacing $\gamma_2^I,\gamma_2^{ij}$ with their expansions in terms of harmonic
forms, we can re-write this action as
\bea 
S&=& -\frac 34 \frac{\t V}{\kappa^2} \int d^4x\left\{e^{4\Omega_{\0}}
\del_\mu c\del^{\h\mu}c + e^{2\Omega_{\0}} C^{IJ}\del_\mu b^I\del^{\h\mu}b^J +g_s
\del_\mu a^i\del^{\h\mu}\b a^j\right.\nonumber\\
&&\left. +2e^{2\Omega_{\0}} C^{IJ}d^{Jij}\del_\mu b^I\alpha^{\h\mu\, ij} 
+e^{2\Omega_{\0}}C^{IJ}d^{Iij}d^{Jkl}\alpha_\mu^{ij}\alpha^{\h\mu\, kl}\right\}\ .
\label{kineticaction3}\eea
Any dependence on the warp factor or 3-form flux is implicit in the 
expansion coefficients $C^{IJ}$.  

Of course, this kinetic action must be invariant under 10D gauge transformations
of the fluctuations.  While we do not present a full analysis here, we sketch
how the constraint equations enforce invariance under the 2-form and 4-form
gauge transformations.  For simplicity of discussion, consider gauge 
transformations just of the $C_4$ axion fluctuations.
Seeing that the action is invariant under the gauge
equivalence $\delta C_4=\delta b_0 \omega_4-\h d b_0 K_3 \to
\delta b(\omega_4 +\t dK_3)$ is simple because the CY parts of the EOM 
$\delta E_6$ are all closed.  The 2-form gauge invariance is a bit more subtle,
as taking $\delta A_2 =-\h db_0\Lambda_1\to \delta b_0 \t d\Lambda_1$ also
shifts $\delta C_4$ by $\delta b_0 (ig_s/2)(\Lambda_1\bG-\b\Lambda_1\G)$, so
and these terms contribute to both $\int \delta C_4\delta E_6$ and
$\int (\delta A_2\delta\b E_8+\delta\b A_2\delta E_8)$.  Some terms cancel
between the two contributions, and the others turn out to be proportional to 
the constraint (\ref{C43formpoisson}).  As before, we see that gauge
invariance is only preserved on the constraint surface.

Since $\kappa^2/\t V$ is the 4D Planck constant $\kappa_4^2$, equation
(\ref{kineticaction3}) takes the 
appropriate form (\ref{Smoduli}) for a metric on moduli space, as 
evaluated at the fixed background.  In other words, we have found 
$G_{ab}(\phi_{\0})$.  Since we have evaluated the metric at an arbitrary point
on moduli space, though, we have the full metric $G_{ab}(\phi)$.  
4D supergravity demands that it is possible to choose holomorphic variables
(corresponding to the scalars of chiral multiplets)
with a Hermitian metric; that task is the subject of the next subsection.

\subsection{K\"ahler metric and potential}

Our results clearly appear similar to the effective action derived in 
\cite{hep-th/0403067} for unwarped orientifolds in which the zero modes
for all axion degrees of freedom are harmonic forms.  Other than
powers of the Einstein-frame factor $e^{\Omega}$, the difference lies
in the interpretation of $C^{IJ}$ versus the metric $G_{\alpha\beta}$ in
\cite{hep-th/0403067}.  Since we hold the K\"ahler structure of
the CY fixed, we have used a basis for $H_{1,1}^+$ in which $G_{\alpha\beta}$
is already (proportional to) the identity, whereas $C^{IJ}$ generically is
not, potentially due to both the warp factor and 3-form flux.  Furthermore,
since $C^{IJ}$ does not scale simply with the volume modulus as does 
$G_{\alpha\beta}$ in the unwarped case, there is not generally any way to 
redefine the fields to transform our K\"ahler metric into that of 
\cite{hep-th/0403067}.  (Of course, either K\"ahler metric can be diagonalized
at any given point in moduli space, but the diagonalization is different
depending on the presence of flux and warping, as we have shown.)

However, on a torus orientifold, the coefficients $C^{IJ}=e^{2\Omega}\delta^{IJ}$
because tori are metrically formal as discussed in section \ref{formality}
and because physical massless modes satisfy $\omega_2\G=0$.  Therefore, the
flux and warp factor leave the moduli space metric unchanged; in fact, 
the moduli space metric is protected from any corrections in 
highly supersymmetric compactifications (with $\N=3,4$ supersymmetry in the
4D theory).  Since more general CY orientifold (including those with a K3 factor)
compactifications necessarily have $\N\leq 2$ supersymetry, the formality of
the torus singles out highly supersymmetric theories even though our 
methodology makes no use of supersymmetry (less supersymmetric torus 
compactifications are also accidentally protected from these corrections).

The generically modified K\"ahler metric on moduli space corresponds to a 
corrected K\"ahler potential.  In the simplifying case that $h_{1,1}^+=1$,
we have noted that the special properties of the K\"ahler form simplify
the expansion coefficients to $C^{IJ}=e^{2\Omega}\delta^{IJ}$ as for a torus 
orientifold.  In this case, it is straightforward to check that the holomorphic
variables and K\"ahler potential
\beq{simplekahler1} a^i,\quad
\rho = c+ib+\frac{g_s}{4} a^i(a^i-\b a^i),\quad
\mathcal{K}(\rho,\b\rho) = -3\ln\left(\rho+\b\rho-\frac {g_s}{4} (a^i-\b a^i)
(a^i-\b a^i)+2e^{-2\Omega_0}\right)\eeq 
yield the kinetic action (\ref{kineticaction3}).  In this case,
our K\"ahler potential agrees with that found by \cite{arXiv:0902.4031} 
using arguments based on 4D and 10D supersymetry also for the restricted case 
that $h_{1,1}^+=1$.
It is of course possible to remove the additive constant from the logarithm
by redefining the holomorphic variable to $\rho=e^{-2\Omega}+ib+(g_s/4)a^i(a^i-\b a^i)$,
as was noted in \cite{arXiv:0810.5768}.  After this 
shift, the K\"ahler potential also agrees with the unwarped result of 
\cite{hep-th/0403067}; however, the shift in the definition of the modulus
does change the value of any nonperturbative superpotential due to Euclidean
D3 instantons or gaugino condensation on D7-branes.

Determining the holomorphic variables and K\"ahler potential that leads to
the metric (\ref{kineticaction3}) is more difficult in the case that 
$h_{1,1}^+>1$ (and we can at best conjecture a K\"ahler potential since we have
not considered the K\"ahler moduli of the CY metric).  However, following
\cite{hep-th/0403067}, we propose that the correct holomorphic variables are
a direct generalization of (\ref{simplekahler1})
\beq{fullkahler1} a^i,\quad
\rho^I = c^I+ib^I-\frac{g_s}{4}d^{Iij}a^i(a^i-\b a^j)\ , \eeq
where $c^1 =e^{-2\Omega}$, the volume modulus, in the basis expanding around a fixed 
point of moduli space, and the other $c^I$ represent the K\"ahler moduli
of the CY metric which leave the volume invariant.  For the basis of
harmonic 2-forms we have chosen, $c^{I\neq 1}=0$ at our fixed point of moduli
space.  In type IIB CY 
compactifications, the actual metric moduli are implicitly defined in terms of 
the $c^I$.  As argued in \cite{arXiv:1205.5728}, a K\"ahler potential with
no-scale symmetry takes the form $\mathcal{K}=-\ln Y$, where $Y$ is a function
of the $c^I$ satisfying $c^I\del Y/\del c^I=3Y$.  The behavior of 
$\mathcal{K}$ required by symmetries of IIB string theory is discussed in
\cite{Grimm:2007xm}.

A K\"ahler potential of this form yields the moduli space metric given by 
(\ref{kineticaction3}) if
\beq{fullkahler2}
\frac{\del Y}{\del c^I} = \frac{3Y}{c^1}\delta^{I1},\quad
e^{2\Omega}C^{IJ}=\frac{1}{Y^2}\frac{\del Y}{\del c^I}\frac{\del Y}{\del c^J}-
\frac{1}{Y}\frac{\del^2 Y}{\del c^I\del c^J}\eeq
at $c^1=e^{-2\Omega},c^{I\neq 1}=0$.  In combination with the fact that 
$C^{I1}=e^{2\Omega}\delta^{I1}$, we find
\beq{fullkahler3} Y = (c^1)^3 -\left[e^{-2\Omega}C^{IJ}\right]_{c^{I\neq 1}=0}
c^1 c^I c^J + \cdots\ .\eeq
Despite some similarities with the analogous expansion for unwarped CY 
compactifications, the dependence of $C^{IJ}$ on the K\"ahler moduli has not
yet been worked out (generally, both the warp factor and flux terms depend on
the $c^I$), and we leave that question to future work.  In particular, the 
unwarped compactification has $Y=\t V^2$ (evaluated as a function of the moduli
and considering $\t g_{mn}$ to scale as $(c^1)^{1/2}$),
but taking the guess of 
$Y$ equal to the squared warped volume does not appear to give
correct K\"ahler potential in the warped case.  Nonetheless, equation 
(\ref{CIJlargevolume}) allows us to evaluate the K\"ahler potential in the 
large-volume limit, since $C^{IJ}$ goes to the standard CY result plus corrections:
\beq{fullkahler4} \mathcal{K} = -\ln \t V^2 + \mathcal{O} \left(e^{2\Omega}\right)
\ .\eeq
That represents corrections to the K\"ahler which scale as $\t V^{-2/3}$, the same
as the higher-curvature corrections discussed in \cite{Grimm:2013gma,Pedro:2013qga}.

The analogous calculation
for D3-brane position moduli will appear in \cite{cfmu}.

\section{Discussion}\label{discuss}
Flux compactifications of the type considered here have played a large role
in the development of string phenomenology and cosmology, primarily due to the
stabilization of moduli by 3-form flux.  We present for the first time 
evidence that the flux and warp factor modify the geometry of the remaining
moduli space, as well.  These corrections are important for models of 
string cosmology and phenomenology based on 4D effective field theory; the issue
is starting from the effective theory that descends from a given compactification.
While we do not expect our results to change qualitative features of 
the effective field theory, such as the existence of many metastable vacua
upon nonperturbative stabilization of the remaining moduli \cite{hep-th/0301240},
quantitative results will be affected.  For example, a number of string
embeddings of inflation use one of the moduli considered here as the inflaton, and
corrections to the inflaton K\"ahler potential likely modify where the 
potential is flat enough to inflate, the normalization of cosmological 
perturbations, the slope of the spectrum, and size of the bispectrum, just to
name a few commonly calculated quantities. We note that the 
modifications to the K\"ahler potential give small but nontrivial corrections 
at large compactification volume,
so large-volume compactifications such as \cite{hep-th/0502058,hep-th/0505076} 
have modified moduli spaces even though the warp factor becomes trivial at 
large volume. Indeed, moduli stabilization in these models 
depends on the large-volume behavior of the K\"ahler potential, and the effects of
K\"ahler potential corrections with the same volume scaling as ours have 
recently been discussed in \cite{Grimm:2013gma,Pedro:2013qga}.  Unlike those
corrections, however, those we discuss are present already in classical supergravity
without higher-curvature terms.  Further, the length scale at which our 
corrections become important is set by flux and warping, rather than the volumes
of cycles wrapped by branes.  In addition, our results apply even in 
compactifications without O7-planes.

As has also been emphasized elsewhere, the key issue is the presence of
nontrivial constraints due to 10D gauge and diffeomorphism invariance.  Solving
the constraints, which must be satisfied for any fluctuation around the
background compactification, typically requires introducing compensators or,
in other words, mixing fluctuations in different 10D fields to form a single
4D degree of freedom.  The 4D effective action is then the action for the
correct 10D fluctuations; satisfying the constraints ensures that the 4D action
is invariant under the 10D gauge symmetry.  In one sense, the constraints
require the mixing of the full tower of CY Kaluza-Klein modes into the zero-mode,
as suggested by \cite{hep-th/0307084}; our results suggest that we should
more properly think of a single zero-mode of the warped compactification with
flux.  We leave to the future to check that our zero-modes are indeed orthogonal
to higher Kaluza-Klein modes (though this was checked in a limited number 
of cases in \cite{arXiv:0810.5768}).

We also leave to the future several other topics, including the application of
our techniques to the K\"ahler moduli of the CY and to brane positions
\cite{cfmu} and the extension of our results to higher order in fluctuations.
The first is ultimately necessary to build a complete 4D effective theory
at the classical level for use with nonperturbative superpotentials in string
phenomenology and cosmology.  The second is technically necessary to derive the
full effective potential for massive modes, although working at linear order
would be sufficient to determine the masses of stabilized moduli in the presence
of warping.  The solutions we have presented at linear order should also be
used to find the interaction of these bulk axions with D-branes, as in
axion monodromy inflation \cite{arXiv:0808.0706}.  A parallel question is
how nonperturbative moduli stabilization appears in 10D \cite{hep-th/0507202,
arXiv:0707.1038,arXiv:0912.4268,arXiv:1001.5028,arXiv:1012.4018};
a complete understanding of nonperturbative moduli stabilization will require
understanding both nonperturbative modifications of the 10D geometry and 
the correspondence between 10D fluctuations and 4D degrees of freedom that
we have discussed.

Our work has also highlighted the question of when the wedge product of
harmonic forms is necessarily harmonic --- the corrections we have found to 
the metric on moduli space are related to the deviation of such wedge 
products from harmonicity.  Some metrics, such as the flat metric on a torus, 
possess the property, known as formality, that wedge products of harmonic
forms are always harmonic.  We have seen an interesting interplay of this
concept with supersymmetry: general Calabi-Yau orientifolds (with $\N=0,1$
supersymmetry) do not support formal metrics and suffer corrections to moduli
space due to warping and flux.
On the other hand, the moduli space of $\N=3,4$ supergravity in
4D is determined by supersymmetry (and the number of gauge symmetries)
and cannot be modified by choices of flux
or the warp factor \cite{deRoo:1984gd,Bergshoeff:1985ms,Castellani:1985ka}.
We have seen that it is precisely the formality of the torus metric in 
$\N=3,4$ orientifold compactifications that protects the moduli space from 
warping and flux corrections.

We close with a short comment on more general flux compactifications.  It seems
reasonable that the moduli of a general flux compactification can be determined
without too much difficulty, perhaps even in terms of topological properties
of the compactification.  After all, as we have stated, the moduli are simply 
deformations that take one allowed background into another.  However, it should
be clear from our work that the geometry of moduli space (and other properties
of the 4D effective field theory) cannot be inferred simply by comparison
to CY compactifications.  Understanding the correct 4D field theory that descends
from a given compactification requires a 10D description and approach.

\acknowledgments
ARF thanks J. Figueroa-O'Farrill, F. Denef, and D. Kotschick
for discussions of metric formality; T. Grimm for a discussion of the holomorphic
variables; G. Torroba and B. Underwood for
collaboration on an early attempt at this calculation; and B. Cownden, D. Marsh, 
and B. Underwood for collaboration on a related project and comments on this
draft. We would particularly like to thank B. Underwood for extensive conversations
on related ideas, especially regarding gauge choices.
This work has been supported by the NSERC of Canada Discovery Grant program.

\appendix

\section{Conventions}\label{conventions}

In this appendix, we briefly summarize our conventions.  External, noncompact
spacetime coordinates are $x^\mu$, while internal, compact dimension coordinates
are $y^m$; when used, $X^M$ are all coordinates.  We work with a
mostly + metric and define the antisymmetric symbol $\epsilon$ as a tensor.
Starting in section \ref{review}, quantities with a hat $\hat{}$ are associated
with the 4D metric $\h\eta_{\mu\nu}$, such as raised or lowered indices, the
antisymmetric tensor $\h\epsilon_{\mu\nu\lambda\rho}$ (or volume form $\h\epsilon$).
Similarly, any quantity with a tilde $\t{}$ is associated with the unwarped
CY metric $\t g_{mn}$.  Forms are treated as in appendix B of 
\cite{Polchinski:1998rr}.  Wedge products are denoted explicitly in equations 
on separate lines, but the wedge symbol is omitted in in-line mathematics.

We consider a basis $\{\omega_2^I\}$ for the harmonic representatives of $H_{1,1}^+$
on the CY manifold, meaning that any harmonic form in $H_{1,1}^+$ can be written
as $e^I\omega_2^I$ for constant coefficients $e^I$.  (This is not the basis of 
harmonic forms at a single point; indeed, if $h_{1,1}^+$ is greater than the 2nd 
Betti number of $T^6$, the $\omega_2^I$ are not all linearly independent at
any given point, only as functions.)  We orthogonormalize the basis with respect
to the following inner product:
\beq{innerproduct} \int \omega_2^I\w\t\star\omega_2^J =3\t V \delta^{IJ}\ .\eeq
This normalization is consistent with setting the first element of the basis
equal to the K\"ahler form $\omega_2^1=\t J$, which satisfies $\t J^3=6\t\epsilon$
and $\t\star\t J =\t J^2/2$.  We write a similarly orthonormalized basis
$\{\omega_2^i\}$ for the harmonic 2-forms in $H_{1,1}^-$.  The triple intersection
numbers are defined with a similar normalization
\beq{triple} 3\t V d^{IJK}=\int\omega_2^I\w\omega_2^J\w\omega_2^K\ ,\quad
3\t V d^{Iij}=\int\omega_2^I\w\omega_2^i\w\omega_2^j\ .\eeq
Then the triple intersection of $\t J$ with itself is $d^{111}=2$.

\section{Linearized equations of motion and quadratic action for fluctuations}
\label{linearizedeom}

Here we derive the expression used in sections \ref{dimred} and
\ref{kahlerpotential} for the kinetic action of moduli.  We will first
show that the second-order action of a general theory around a fixed
background is given by the first-order fluctuation contracted with the
linearized equations of motion.  While this principle has appeared
elsewhere, we are not aware of a general proof, so we present it here.
We then find the corresponding formula for type IIB supergravity.

\subsection{General formalism}
Consider a theory with fields $\Phi_A$, where $A$ includes the field,
coordinate indices, gauge indices, etc., and action
\beq{generalaction}
S =\int d^nx\, \L(\Phi,\del_M\Phi,\del_M\del_N\Phi)\eeq
with up to two derivatives allowed (although we know of no obstruction
to generalizing to  higher-derivative theories).  We have included
$\sqrt{-g}$ in $\L$, so $\L$ is a scalar density.  We write the
Lagrangian in term of partial derivatives.  Ignoring boundary terms,
the equations of motion (EOM) are
\beq{generalEOM1}
\frac{\delta\L}{\delta\Phi_A}-\del_M\frac{\delta\L}{\delta\del_M\Phi_A}+\del_M\del_N
\frac{\delta\L}{\delta \del_M\del_N\Phi_A}=0\ .\eeq
Note that, despite the fact that we are not using manifestly covariant
notation, the EOM are tensor densities as long as $\Phi_A$ are
tensors.

Now we consider a perturbative expansion of the fields around a fixed
classical background $\Phi^0$,  $\Phi=\Phi^0+\Phi^1+\Phi^2+\cdots$,
where $\Phi^n$ is $n$th order in some small parameter.  The linearized
EOM are
\bea
0&=&\frac{\delta^2\L}{\delta\Phi_A\delta\Phi_B}(\Phi^0)\Phi_B^1+
\frac{\delta^2\L}{\delta\Phi_A\delta\del_M\Phi_B}(\Phi^0)\del_M\Phi_B^1+
\frac{\delta^2\L}{\delta\Phi_A\delta\del_M\del_N\Phi_B}(\Phi^0)\del_M\del_N\Phi_B^1
\nonumber\\
&&-\del_M\left(\frac{\delta^2\L}{\delta\del_M\Phi_A\delta\Phi_B}(\Phi^0)\Phi_B^1\right)
-\del_M\left(\frac{\delta^2\L}{\delta\del_M\Phi_A\delta\del_N\Phi_B}(\Phi^0)
\del_N\Phi_B^1\right)\nonumber\\
&&+\del_M\del_N\left(\frac{\delta^2\L}{\delta\del_M\del_N\Phi_A\delta\Phi_B}(\Phi^0)
\Phi_B^1\right)\label{generalEOM2}
\eea
Recall that we have restricted to terms with up to two derivatives only.

We can similarly expand the action to second order:
\bea
S&=& \int d^nx\,\left\{\L(\Phi^0)+\left[\frac{\delta\L}{\delta\Phi_A}(\Phi^0)
\left(\Phi_A^1+\Phi_A^2
\right)+\frac{\delta\L}{\delta\del_M\Phi_A}(\Phi^0)\del_M
\left(\Phi_A^1+\Phi_A^2\right)\right.\right.
\nonumber\\
&&\left.\left.
+\frac{\delta\L}{\delta\del_M\del_N\Phi_A}(\Phi^0)\del_M\del_N
\left(\Phi_A^1+\Phi_A^2\right)\right]
+\frac 12 \left[\frac{\delta^2\L}{\delta\Phi_A\delta\Phi_B}(\Phi^0)\Phi_A^1\Phi_B^1
\right.\right.\nonumber\\
&&\left.\left. +\frac{\delta^2\L}{\delta\Phi_A\delta\del_M\Phi_B}(\Phi^0)\Phi_A^1
\del_M\Phi_B^1
+\frac{\delta^2\L}{\delta\Phi_A\delta\del_M\del_N\Phi_B}(\Phi^0)\Phi_A^1
\del_M\del_N\Phi_B^1\right.\right.\nonumber\\
&&\left.\left. +\frac{\delta^2\L}{\delta\del_M\Phi_A\delta\Phi_B}(\Phi^0)
\del_M\Phi_A^1\Phi_B^1
+\frac{\delta^2\L}{\delta\del_M\Phi_A\delta\del_N\Phi_B}(\Phi^0)\del_M\Phi_A^1
\del_N\Phi_B^1\right.\right.\nonumber\\
&&\left.\left. +\frac{\delta^2\L}{\delta\del_M\del_N\Phi_A\delta\Phi_B}(\Phi^0)
\del_M\del_N\Phi_A^1\Phi_B^1\right]\right\}\ .\label{generalaction2} 
\eea
After integration by parts, the terms in the first square brackets are
proportional to the background EOM and vanish by assumption.  If we
integrate by parts to move derivatives off of $\Phi_A^1$ in the second
square brackets, they become $1/2$ times the linearized EOM of
$\Phi_A$ contracted with $\Phi_A^1$.  Besides this result, note that we have
shown that the second order fluctuations $\Phi_A^2$ in the fields do not 
contribute to the action at second order.

\subsection{Type IIB SUGRA}\label{IIBEOM}

The 10D type IIB SUGRA action in Einstein frame is
\beq{SUGRA1}
S=\frac{1}{2\kappa^2} \int d^{10}x\, \sqrt{-g}\, R -\frac{1}{2\kappa^2}
\int \left( \frac{g_s}{2} G_3\w\star\b G_3 +\frac 14 \tF_5\w\star\tF_5
+\frac{ig_s}{4} C_4\w G_3\w\b G_3\right)\ ,\eeq
where we impose that the axio-dilaton is fixed at $\tau=i/g_s$ and work with a
complex 3-form field strength.  Note that the above action can be written
entirely in terms of contractions of the fields to make a connection with 
the formalism given above.  Here, we list the 10D EOM for this theory; as 
defined above, the EOM are the first functional derivatives of the Lagrangian
density and are slightly different from the usually quoted equations. As 
usual, 5-form self-duality $\tF_5=\star\tF_5$ must be enforced by hand as a 
constraint at the level of the equations of motion.

To start, the Einstein equation is 
\beq{sugraeinstein}
E_{MN} = %\frac{\sqrt{-g}}{2\kappa^2}
\left[ G_{MN} -\frac{g_s}{4}
G_{(M}{}^{PQ}\b G_{N)PQ} +\frac{g_s}{4}g_{MN}|G_3|^2 -\frac{1}{96}\tF_M{}^{PQRS}\tF_{NPQRS}
+\frac 18 g_{MN} |\tF_5|^2\right]=0\ ,
\eeq
where $G_{MN}$ is the Einstein tensor 
($\delta S/\delta g^{MN}=(\sqrt{-g}/2\kappa^2) E_{MN}$).  
Once self-duality is imposed, we can 
re-write this as the usual Einstein equation
\beq{sugraeinstein2} G_{MN}=T^3_{MN}+T^5_{MN},\quad
T^3_{MN}=\frac{g_s}{4}G_{(M}{}^{PQ}\b G_{N)PQ} -\frac{g_s}{4}g_{MN}|G_3|^2 ,\quad
T^5_{MN}=\frac{1}{96}\tF_M{}^{PQRS}\tF_{NPQRS}\eeq
in terms of stress-energy tensors for the 3-form and 5-form.

The EOM for the 4-form potential can be written as a 6-form equation
\beq{sugraE6}
E_6 = \left[ d\star\tF_5-\frac{ig_s}{2}G_3\w\b G_3\right]=0\ .\eeq
Similarly, the EOM for $A_2$ is an 8-form
\beq{sugraE8}
\b E_8 = \left[ d\star\b G_3-\frac i2 \b G_3\w\left(
\star\tF_5+\tF_5\right) -\frac i4 \b A_2 \w E_6\right]=0\ .\eeq
Usually, $\b E_8$ and its conjugate are written with 5-form self-duality and 
$E_6=0$ imposed; however, we must retain $E_6\neq 0$ when evaluating the 
off-shell action.  These EOM are normalized so 
\beq{sugraformnorms}
\delta S = \frac{1}{2\kappa^2} \int \left[\frac 12 \delta C_4\w E_6
+\frac{g_s}{2}\delta A_2\w \b E_8+\frac{g_s}{2}\delta\b A_2\w E_8\right]\ .\eeq

One subtlety of IIB SUGRA is the self-duality of $\tF_5$.  As is well-known,
the kinetic action from eqn. (\ref{SUGRA1}) vanishes on a self-dual 5-form.
We adopt the prescription of setting half the degrees of freedom of $C_4$ to zero and
doubling the contribution to the action from those we retain. In this paper,
we are concerned with 4D scalars, so this prescription amounts to keeping 
components of $C_4$ with zero or one leg along the external spacetime.  We
also keep components with two external legs only when they include a leg
along the spacetime gradient of the modulus.  As a part of this doubling,
the dynamical EOM for $\delta A_2$ becomes
\beq{sugraE8b}
\delta\b E_8 = d\delta(\star\b G_3) -i\b G_3\w\delta(\star\tF_5)-\frac i2
\b A_2\w \delta E_6\ ,\eeq
which keeps components with $\geq 3$ spacetime indices (note that $\tF_5$ 
does not appear because it has fewer spacetime indices but that the transgression
terms have been doubled).

In the end, the quadratic action with fluctuations becomes
\beq{SUGRA2}
S = \frac{1}{4\kappa^2} \int d^{10}x\, \sqrt{-g}\, \delta g^{MN}
\delta E_{MN} +\frac{1}{4\kappa^2}\int\left(\delta C_4\w\delta E_6+
\frac{g_s}{2}\delta A_2\w\delta\b E_8 +\frac{g_s}{2}\delta\b A_2 \w\delta E_8 
\right)\ ,\eeq
where $\delta$ represents the first order part and we have doubled the
contribution of the remaining 4-form components as discussed above.  
In terms of contractions, this is
\bea
S&=&\frac{1}{4\kappa^2} \int d^{10}x\, \sqrt{-g}\left[\delta g^{MN}\delta E_{MN}
+\delta C^{MNPQ}\left(\star\delta E_6\right)_{MNPQ}\right.\nonumber\\
&&\left.+\frac{g_s}{2}\delta A^{MN}
\left(\star\delta\b E_8\right)_{MN}+\frac{g_s}{2}\delta\b A^{MN}\left(\star
\delta E_8\right)_{MN}\right]\ .\label{SUGRA3}
\eea
The Hodge star is zeroth-order since we assume that the background EOM are 
satisfied.

\subsection{The 4-form in 3-form flux}\label{C4inG3}

As has been discussed in \cite{hep-th/0201029,arXiv:0810.5768}, there are 
additional subtleties in the dimensional reduction of $C_4$ in the presence
of nontrivial harmonic $G_3$ due to the behavior of $C_4$ under the gauge
transformations of $A_2$.  The point is that, since $G_3^{\0}=\t dA_2$ is harmonic, 
$A_2$ cannot be globally well-defined but is instead glued together by
gauge transformations between at least two patches on the internal manifold.
$C_4$ is therefore also not globally defined.  This carries over to first
order in fluctuations; in dimensional reduction, the globally-defined part 
of $\delta C_4$ leads to the 4D degree of freedom.

To find this globally-defined part, we split $A_2$ into a
``flux part'' and a ``global part,'' $A_2=A_2^{(f)}+A_2^{(g)}$, with 
$A_2^{(f)}(y)$ zeroth-order and $\t dA_2^{(f)}=G_3^{\0}$.
$A_2^{(g)}(x,y)$ contains both zeroth- and first-order terms; the background
must be closed and can be chosen harmonic, 
so $dA_2^{(g)}=\delta G_3$ is purely first-order.  We write 
$C_4=C_4^{\0}+\delta C_4$ as a sum of background and fluctuations.

Between two gauge patches, $A_2^{(f)}$ is glued together
by a (fixed) gauge transformation $\lambda_1$.
Since the general gauge transformations of IIB SUGRA are
\beq{gauge1}
A_2\to A_2+d\zeta_1 ,\quad C_4 \to C_4+d\chi_3-\frac{ig_s}{4}\left( \zeta_1\w \b G_3
-\b\zeta_1\w G_3\right)\ ,
\eeq
we also must have a gluing of
\beq{gauge2} C_4^{\0}\to C_4^{\0}+\t d\Gamma_3 -\frac{ig_s}{4}
\left(\lambda_1\w \b G_3^{\0}-\b\lambda_1\w G_3^{\0}\right),\quad \delta
C_4\to \delta C_4+d\Upsilon_3 -\frac{ig_s}{4}\left(\lambda_1\w \delta \b G_3
-\b\lambda_1\w \delta G_3\right)\ .\eeq
Therefore, $\delta C_4$ as written is not globally-defined.  To proceed,
we note that 
\beq{gauge3}
\lambda_1\w \delta\b G_3 = -d\left(\lambda_1\w\b A_2^{(g)}\right)+
(\t d\lambda_1)\w \b A_2^{(g)}\ ,\eeq
where the last term is the transition gauge transformation of $A_2^{(f)}\b A_2^{(g)}$.
As a result, 
\beq{deltaCp}
\delta C'_4 = \delta C_4 +\frac{ig_s}{4}\left( A_2^{(f)}\w \b A_2^{(g)} -
\b A_2^{(f)}\w A_2^{(g)}\right)\eeq
has the transition gauge transformation
\beq{gauge4}
\delta C'_4\to \delta C'_4 +d\Upsilon_3
+\frac{ig_s}{4}d\left(\lambda_1\w \b A_2^{(g)}-\b\lambda_1\w A_2^{(g)}\right)\ .
\eeq
With the appropriate choice of $\Upsilon_3$, $\delta C'_4$ is globally defined.
Its general gauge transformation is
\beq{gauge5}\delta C'_4\to \delta C'_4 +d\chi_3-\frac{ig_s}{2}\left(\zeta_1
\w\b G_3^{\0}-\b\zeta_1\w G_3^{\0}\right) -\frac{ig_s}{4}\left( \zeta_1
\w\delta\b G_3-\b\zeta_1\w\delta G_3\right)\ .\eeq

We then should write the quadratic action in terms of $\delta C'_4$ rather
than $\delta C_4$ since it represents the 4D degree of freedom.  This simply
replaces $\delta C_4\to \delta C'_4$ in (\ref{SUGRA2}) and a simultaneous
replacement of $\b A_2\to \b A_2^{(g)}$ in the last term of $\b E_8$ in 
equation (\ref{sugraE8}). 
For reference, the 5-form in terms of globally-defined
variables is written as
\bea
\tF_5 &=& dC_4^{\0}+\frac{ig_s}{4}\left(A_2^{(f)} \w\b G_3^{\0}-\b A_2^{(f)} \w G_3^{\0}
\right)+\frac{ig_s}{2}\left(A_2^{(g)} \w\b G_3^{\0}
-\b A_2^{(g)} \w G_3^{\0}\right)\nonumber\\
&& +d\delta C'_4+\frac{ig_s}{4}\left(A_2^{(g)} \w\delta\b G_3
-\b A_2^{(g)} \w \delta G_3 \right)\ .\label{gauge6}
\eea
Note that the last set of terms on the first line can include both background
(zeroth-order) and fluctuation (first-order) parts in principle.

\section{Ansatz for linear perturbations}\label{offdiagonal}

Throughout, we have worked with a metric of the form
\beq{generalBmetric}
ds^2=e^{2\Omega(x)}e^{2A(x,y)}\hat\eta_{\mu\nu}dx^\mu dx^\nu +2 e^{2\Omega(x)}e^{2A(x,y)} 
\del_\mu\B_m(x,y) dx^\mu dy^m + e^{-2A(x,y)} \t g_{mn} dy^m dy^n\ ,
\eeq
where the internal metric $\t g_{mn}$ is Calabi-Yau and depends only on $y^m$.  
The associated 5-form and 3-form field strengths are 
\bea
\t F_5 &=& e^{4\Omega(x)} \h\epsilon\w \t de^{4A(x,y)} +\t\star\t de^{-4A(x,y)}
+e^{4\Omega(x)} \h\star \h d \B_1\w \t de^{4A(x,y)}\nonumber\\
&&+\bm{\alpha}_1\w\bm{\omega}_4
+e^{2\Omega(x)}e^{4A(x,y)}\h\star\bm{\alpha}_1\w\t\star\bm{\omega}_4
\label{generalB5form}\\
G_3 &=&G_3^{\0} +\bm{\beta}_1\w\bm{\omega}_2\ ,\ \ \t\star G_3^{\0}=iG_3^{\0}\ ,
\label{generalB3form}\eea
where the last terms depend on the particular modulus under 
consideration ($\bm{\alpha}_1(x), \bm{\beta}_1(x)$ are closed spacetime 
forms to linear order, while 
$\bm{\omega}_4(y),\bm{\omega}_2(y)$ are forms on the internal CY). 
We work at linear order in $x^\mu$ dependent fluctuations
(in fact, the metric ansatz should be modified at higher order to account
for backreaction on the external spacetime). 
In the main text, $\B_1$ is generally a 
superposition of products of a function of $x^\mu$ and a 1-form on the 
internal space.  Here we collect a few results that are useful through the
entire paper.\footnote{Some of these results were first obtained for
\cite{cfmu}.}

\subsection{Einstein tensor}

Both diffeomorphism constraints and dynamical equations of motion arise
from the linearized Einstein equation.  The Einstein tensor for the metric
(\ref{generalBmetric}) is
\bea
G_{\mu\nu} &=& -2e^{2\Omega}e^{4A}\left(\t\Del^{\t 2} A -2\del_m A\del^{\t m}A\right)
\h\eta_{\mu\nu}
-2\left(\del_\mu\del_\nu \Omega-2\del_\mu \del_\nu A -\frac 12 e^{2\Omega}e^{4A} 
\del_\mu \del_\nu \t\Del^{\t m} \B_m\right) \nonumber\\
&&+2\left(\del^{\hat 2}\Omega-2\del^{\h 2}A-\frac 12 e^{2\Omega}e^{4A} \del^{\h 2}
\t\Del^{\t m}
\B_m\right)\h\eta_{\mu\nu}\ ,\\
G_{\mu m} &=& -\frac 12 e^{4A} \del_\mu \del_m e^{-4A} +\frac 12 e^{2\Omega}e^{4A}
\left(\del_\mu \t\Del^{\t n}(\t d\B)_{mn} +4 \del^{\t n}A\del_\mu(\t d\B)_{mn}
\right)\nonumber\\
&&-2e^{2\Omega}e^{4A}\left(\t\Del^{\t 2} A -2\del_m A\del^{\t m}A\right)\del_\mu\B_m
\ ,\\
G_{mn}&=& -8\del_m A\del_n A+4(\del_p A\del^{\t p}A)\t g_{mn} +e^{-2\Omega}e^{-4A}
\left(3\del^{\h 2}\Omega-2\del^{\h 2}A\right)\t g_{mn}\nonumber\\
&&+\left(\del^{\h 2}\t\Del_{(m}\B_{n)} +4\del_{(m}A\del^{\h 2}\B_{n)} -
\del^{\h 2}(\t\Del^{\t p}\B_p)\t g_{mn} -2\del^{\t p}A\del^{\h 2}\B_p\,\t g_{mn}\right)
\ .\eea
These include terms at order zero and one in spacetime dependent fluctuations;
in particular, the first terms in both $G_{\mu\nu}$ and $G_{mn}$, which appear
in the zeroth order Einstein tensor, should properly be expanded to first 
order.  However, these terms and the second line of $G_{\mu m}$ automatically
cancel against the stress tensor
as a consequence of the background (zeroth order) Einstein equations.

Since there are no harmonic 1-forms on a generic Calabi-Yau 3-fold, we
can write $\B_1 = \B'_1-\t dK$, where $\B'_1$ is co-exact.  Then we
find the simplifications
\beq{simplify1}
\t\Del^{\t m}\B_m = -\t\Del^{\t 2} K\quad ,\qquad \t\Del^{\t n}(\t d\B)_{mn}=
-\t\Del^{\t 2}\B'_m\ .\eeq
We have used the fact that $\t g_{mn}$ is Ricci-flat in the second equality.

\subsection{Stress tensor}

The ``off-diagonal'' term in the metric also manifests itself in the stress
tensor.  For example, the stress tensor for $G_3$, $T_{MN}=(g_s/4) 
(G_{(M}{}^{PQ}\b G_{N)PQ}-g_{MN} |G|^2)$, which trivially includes a 
$T_{\mu m}\propto \del_\mu \B_m$ term.  Through first-order terms,
the stress tensor for $G_3$ is
\bea
T_{\mu\nu} &=& -\frac{g_s}{4} e^{2\Omega}e^{8A}|G_3^{\0}|^{\t 2}\h\eta_{\mu\nu}\nonumber\\
T_{\mu m} &=& \frac{ig_s}{4} e^{4A}\left[ \bm{\beta}_\mu \t\star(
\bm{\omega}_2\w\b G_3^{\0})_m -c.c.
\right]-\frac{g_s}{4}e^{2\Omega}e^{8A}\del_\mu \bm{B}_{m}|G_3^{\0}|^{\t 2}\\
T_{mn}&=& 0 \ .\eea
In the first term of $T_{\mu m}$, we have used the ISD property of $G_3^{\0}$
to convert a contraction into a wedge product.  Further, we have used 
the ISD property of the background flux to show that
\beq{T3mn_ISD} G_{(m}{}^{\widetilde{pq}} \b G_{n)pq} =(\t\star G)_{(m}{}^{\widetilde{pq}} 
(\t\star \b G)_{n)pq}
=2\t g_{mn}|G_3^{\0}|^{\t 2} - G_{(m}{}^{\widetilde{pq}} \b G_{n)pq}\ ,\eeq
so the two contributions to $T_{mn}$ cancel.

However, $\B_m$ also appears in
the 5-form stress tensor at linear order due both to its appearance in 
(\ref{generalB5form}) and through contractions.  That is, through first
order, starting with $T_{MN}=(1/96)\t F_{MPQRS}\t F_N{}^{PQRS}$,
\bea
T_{\mu\nu} &=& -4 e^{2\Omega} e^{4A} \left(\del_m A\del^{\t m} A\right)
\h\eta_{\mu\nu}\ ,\\
T_{\mu m}&=& \frac{1}{24} \t F_{\mu\nu\lambda\rho n}\delta\t F_m{}^{\nu\lambda\rho n}
+\frac{1}{24} \t F_{\mu\nu\lambda\rho n}\t F_m{}^{\nu\lambda\rho}{}_\alpha 
\delta g^{\alpha n}\nonumber\\
&=& 
-4e^{2\Omega}e^{4A} \left(\del_n A\del^{\t n}A\right)\del_\mu \B_m-2e^{4A}
\bm{\alpha}_\mu(\t\star\bm{\omega})_{mn}\del^{\t n}A\ ,
\label{generalB_T5offdiag}\\
T_{mn}&=& -8\del_m A\del_n A +4\left(\del_p A\del^{\t p}A\right)\t g_{mn}\ .
\eea
In Eqn. (\ref{generalB_T5offdiag}), we have indicated explicitly the 
appearance of $\B_m$ in both the 5-form and the metric.  Both contributions
are required for the $\del_\mu\B_m$ terms in the Einstein equation to be
proportional to the instantaneously satisfied background equations of motion.

\subsection{Five-form self-duality}
Finally, we confirm that (\ref{generalB5form}) is self-dual.  
To see this, consider that
\beq{generalB_selfduality1}
\star\left(\t\star \t de^{-4A} \right)_{\mu\nu\lambda\rho m} =\frac{1}{5!}
\left(e^{4\Omega}e^{4A} \hat\epsilon_{\mu\nu\lambda\rho}\right) \left(e^{4A} 
\t\epsilon_m{}^{\widetilde{npqrs}}\right)\t\epsilon_{npqrs}{}^{\t t}\del_t e^{-4A}=
-e^{4\Omega}e^{8A}\h\epsilon_{\mu\nu\lambda\rho}\del_m e^{-4A}\ ,
\eeq
which is just component notation for $e^{4\Omega}\h\epsilon\w\t de^{4A}$.
This is self-duality at zeroth-order, though it also includes first order
constributions from $A$ and $\Omega$.  Including the off-diagonal metric,
there is also a component
\bea
\star\left(\t\star \t de^{-4A} \right)_{\mu\nu\lambda mn} &=&\frac{1}{4!}
\left(e^{4\Omega}e^{4A}\h\epsilon_{\mu\nu\lambda\rho}\right)\left(e^{2A}
\t\epsilon_{mnpqrs}\right)\t\epsilon_t{}^{\widetilde{pqrsu}}\del_ue^{-4A}
\delta g^{\rho t}\nonumber\\
&=&2e^{4\Omega}\h\epsilon_{\mu\nu\lambda}{}^{\h\rho}\del_\rho\B_{[m}\del_{n]}e^{4A}\ .
\label{generalB_selfduality2}\eea
This is component notation for $\h\star\h d(\B_1\w\t de^{4A})$.  In other
words, the dual of the second term in (\ref{generalB5form}) is the first
plus third terms.  Since $\star\star=1$ on 5-forms in 10D, we can
conclude that (\ref{generalB5form}) is self-dual since the terms involving
$\bm{\alpha}_1, \bm{\omega}_4$ are clearly self-dual.

\subsection{3-form EOM}

While the 5-form EOM is the same as its Bianchi identity by self-duality,
the $G_3$ EOM requires the 10D Hodge star.  
The key point
is that $\star G_3$ contains first-order contributions from both the metric 
and $G_3$.  Specifically, 
\beq{starG3}\star G_3 = ie^{4\Omega}e^{4A}\h\epsilon\w G_3^{\0}+
e^{2\Omega}\h\star\bm{\beta}_1\w \t\star\bm{\omega}_2 +ie^{4\Omega}e^{4A}
\h\star\h d \bm{B}_1\w G_3^{\0}\ .\eeq
The EOM as usually written is $d\star G_3 - i\tF_5 G_3=0$, so the constraint
follows by acting on (\ref{starG3}) with $\t d$ and subtracting $i\tF_5 G_3$.
An additional simplification occurs because $\tF_5^{\0}\delta G_3=0$ by index
counting, and a number of terms cancel.  The constraint becomes
\beq{G3constraint}
-e^{2\Omega}\h\star\bm\beta_1\w\t d\t\star\bm\omega_2-ie^{2\Omega}e^{4A}\h\star
\bm\alpha_1\w\t\star\bm\omega_4\w\G-ie^{4\Omega}e^{4A}\h\star\h d\left(\t d
\bm{B}_1\right)\w \G=0 \ .
\eeq
The dynamical EOM receives no contribution from the transgression term and is
\beq{G3dynamical}
\delta E_8=e^{2\Omega}\h d\h\star\bm\beta_1\w\t\star\bm\omega_2 +ie^{4\Omega}e^{4A}
\h d\h\star\h d\bm{B}_1\w\G+\frac i2 A_2^{(g)}\w\delta E_6\eeq
following equation (\ref{sugraE8b}).

\bibliographystyle{utcaps2}
\bibliography{formaxion}
\end{document}